\documentclass[structabstract]{aa}

\usepackage{graphicx}
\usepackage{txfonts}
\usepackage{natbib}
\usepackage{url}
\bibpunct{(}{)}{;}{a}{}{,}

\def\pcite#1{[ref]}

\def\<#1>{\langle#1\rangle}
\def\pcite#1{[ref]}

\def\Sec{\hbox{${}^{\prime\prime}$\llap{.}}}
\def\deg{\hbox{${}^\circ$}} 
\def\sec{\hbox{${}^{\prime\prime}$}}
\def\lae{\mathrel{<\kern-1.0em\lower0.9ex\hbox{$\sim$}}}
\def\gae{\mathrel{>\kern-1.0em\lower0.9ex\hbox{$\sim$}}}

\def\etal{et al.~}

\begin{document}
   \title{Non-parametric Deprojection of Surface Brightness Profiles of Galaxies in Generalised Geometries}
   \subtitle{}
   \titlerunning{Deprojecting Brightness Profiles}
   \author{Dalia Chakrabarty \inst{1}
          }
   \institute{
              School of Physics $\&$ Astronomy,
              University of Nottingham, 
              Nottingham NG7 2RD, U.K.
              \email{dalia.chakrabarty$@$nottingham.ac.uk}
              \thanks{current email address: D.Chakrabarty@warwick.ac.uk}
             }

   \date{\today}

   \abstract {}{We present a new Bayesian non-parametric deprojection
     algorithm DOPING (Deprojection of Observed Photometry using and
     INverse Gambit), that is designed to extract 3-D luminosity
     density distributions $\rho$ from observed surface brightness
     maps $I$, in generalised geometries, while taking into account
     changes in intrinsic shape with radius, using a penalised
     likelihood approach and an MCMC optimiser.}{We provide the most
     likely solution to the integral equation that represents
     deprojection of the measured $I$ to $\rho$. In order to keep the
     solution modular, we choose to express $\rho$ as a function of
     the line-of-sight (LOS) coordinate $z$. We calculate the extent
     of the system along the ${\bf z}$-axis, for a given point on the
     image that lies within an identified isophotal annulus. The
     extent along the LOS is binned and density is held a constant
     over each such $z$-bin. The code begins with a seed density and
     at the beginning of an iterative step, the trial $\rho$ is
     updated. Comparison of the projection of the current choice of
     $\rho$ and the observed $I$ defines the likelihood function
     (which is supplemented by Laplacian regularisation), the maximal
     region of which is sought by the optimiser (Metropolis
     Hastings).}{The algorithm is successfully tested on a set of test
     galaxies, the morphology of which ranges from an elliptical
     galaxy with varying eccentricity to an infinitesimally thin disk
     galaxy marked by an abruptly varying eccentricity
     profile. Applications are made to faint dwarf elliptical galaxy
     Ic~3019 and another dwarf elliptical that is characterised by a
     central spheroidal nuclear component superimposed upon a more
     extended flattened component. The result of deprojection of the
     X-ray image of cluster A1413 - assumed triaxial - the axial ratios
     and inclination of which are taken from the literature, is also
     presented.}{}

  \keywords{Astronomical instrumentation, Methods and techniques;
  Methods: statistical; Galaxies: fundamental parameters
  (classification, colours, luminosities, masses, radii, etc.) }

  \maketitle

  \section{Introduction}
  \label{sec:intro}
  \noindent
  An integral step in the construction of dynamical models of galaxies
  is the recovery of the intrinsic luminosity density from the surface
  brightness that is observed projected on the plane of the sky
  \citep{magog98, kronawitter00, krajnovic04}. Such deprojection is
  non-trivial and indeed offers no unique solutions except for very
  specific configurations of geometry and inclination. As demonstrated
  by \cite{rybicki87}, the deprojection is degenerate for axisymmetric
  systems viewed at inclination angles, $i$, other than edge-on
  ($i=90$\deg). This is a consequence of the fact that the observed
  surface brightness cannot yield any information on a density term
  whose Fourier transform is non-zero only within a cone of half angle
  $90^{\circ} -i$ (the ``cone of ignorance'').  \cite{binneygerhard}
  report a family of analytical konus densities. \cite{kochanekrybicki}
  found a family of konus densities that have arbitrary densities in the
  equatorial plane. \cite{vandenbosch} finds that konus densities
  contribute at most a few percent of the total galactic mass to the
  centre of elliptical galaxies with nuclear cusps, implying that their
  dynamical influence is minimal. \cite{magorrian} suggests that nearly
  face-on disk-like konus densities can be recognised via the unique
  signature they imprint on the line-of-sight (LOS) velocity profiles.

  These problems notwithstanding, a great deal of effort has been put
  into the development of methodologies aimed at deprojecting
  two-dimensional photometric information. These include parametric
  formalisms designed by \cite{palmer}, \cite{bendinelli} and
  \cite{cappellari}, as well as non-parametric methods, such as the
  Richardson-Lucy Inversion scheme \citep{richardson, lucy} and a method
  by \cite{romkoch} (hereafter, RK).

  The parametric methods work by making series expansions of the density
  (or brightness) and fit the brightness (or density) to the coefficient
  of the expansions; convergence is defined at a preset accuracy level.
  In \cite{bendinelli}, the density is derived via a Gaussian expansion
  of the surface brightness profile, an idea further developed by
  \cite{cappellari}. In \cite{palmer}, the density is expanded in terms
  of angular polynomials and the projections of these are then fit to
  the surface brightness. These methods suffer from the basic drawback
  that the answer depends on the choice of the basis functions. Thus,
  the solution is forced to conform to a subset of all possible
  solutions. Even more worrisome is the fact that the validity of the
  goodness of fit measures, or ``$\chi^2$ quantities'' that are employed
  in these schemes to identify acceptable fits, is questionable,
  particularly in the presence of inhomogeneous noise
  \citep{bissantz_01}. On the whole, the fitting of non-linear
  functions, to what is usually noise-ridden incomplete data, over large
  dynamical ranges, is worrisome.

  Deprojection of surface brightness profiles has also been attempted
  with the Abel integral equation, under the assumption of sphericity or
  axisymmetry with an edge-on inclination \citep{merrittmeylanmayor,
  merritttremblay, gebhardt96}.

  An example of a non-parametric inversion scheme is the Richardson-Lucy
  algorithm \citep{richardson, lucy}, which has been widely used in the
  stellar dynamical context. It is a simple deprojection scheme that
  works by iterating toward increasingly better approximations to the
  density that fits the data. However, absolute convergence is not
  sought in this framework --- rather, the iterations are stopped when
  the density is judged to be a good fit to the observations. In lieu of
  this imposed clause in the code, progressive iterations would produce
  increasingly more unphysical densities. This lack of a robust
  convergence criterion is cause for dissatisfaction with the Richardson
  Lucy scheme. Moreover, implementations of the same, within Astronomy,
  have not incorporated either radial variations of intrinsic shape or
  deviations from sphericity.

  The shortcomings of Lucy's algorithm are overcome by the analysis of
  RK in their incorporation of monotonicity and positivity into the
  sought solution and by their more satisfying convergence criterion. RK
  construct their density profile as a series of stacked blocks in the
  space of one quadrant of the meridional plane of their axisymmetric
  (by assumption) galaxy. The density values at radially adjacent
  locations are connected through linear interpolation. The summation of
  the projections of each of the density blocks along the LOS, give the
  surface brightness at a given location in the plane of the sky (where
  a brightness measurement is reported). This estimated brightness is
  then compared to the observed brightness; the algorithm attempts to
  minimise this statistic while imposing the smoothing condition through
  a bias function, thus providing a more satisfying convergence
  criterion than included in Lucy's algorithm. However, this scheme too
  fails to allow for deprojection in general triaxial geometries;
  specifically, it is designed to reproduce axisymmetric
  systems. Moreover, the validity of interpolation, in systems where
  local gradients can be steep, is also worrisome. 

  \cite{magorrian} also advances a scheme similar to RK's expect that he
  implements a penalty function in his definition of likelihood. The
  imposed penalty is achieves nearest-neighbour smoothing. The
  fundamental shortcomings of this scheme are the same as what plagues
  RK's method. There is no apriori reason to belive that the galaxy
  under consideration is axisymmetric; triaxiality is a much more
  general model. Moreover, the requirement in this work, for density to
  behave like a power-law on local scales, implies that the method will
  fail in the presence of even moderate gradients in density; in this
  sense, the recovered answer could be sensitive to the binning details
  of the 2D grid on which the density structure is placed.


  In this paper, we present a new, Bayesian non-parametric algorithm
  that implements an MCMC optimiser, in order to tackle the deprojection
  of observed photometry of galaxies. Completely free-form
  solutions for the 3D density are provided with the constraint of
  positivity imposed by hand. The scheme is a penalised likelihood
  procedure. We refer to this algorithm as DOPING - an acronym for
  Deprojection of Observed Photometry using an INverse Gambit. The
  algorithm is easy to implement and each run typically takes about a
  a couple of minutes on a state-of-the art personal computer.

  The most distinguishing feature of DOPING is that it can tackle
  deprojection in virtually any geometry, as long as we can express
  the intrinsic shape parameters (such as eccentricities) in analytical
  relations with the projected shape parameters. DOPING can be applied to
  deproject surface brightness maps of elliptical as well as disk
  galaxies. This is possible, while taking intrinsic shape variation
  into account.

  \footnotetext{Such a broad class of solid shapes are perhaps an
  overkill for astrophysical applications, but these shapes do indeed
  show up in nature, for example, the wide range of shapes of images of
  deposited nanoparticles (taken with Transmission Electron Microscope)
  indicate that these would require such a description \citep{yong_06}.}


  The first major application of this algorithm to galaxies (Chakrabarty
  $\&$ McCall, 2009) is the study of the deprojected luminosity profiles
  of 100 early type galaxies observed as part of the ACS Virgo Cluster
  Survey \citep{coteacs04}. As these systems do not exhibit significant
  variations in position angle, the preliminary version of DOPING that
  is presented here, considers position angle to be a constant. However
  the code can account for changes in position angle and the skeletal
  scheme for the inclusion of a radially varying position angle is
  presented later in \S5.6.

  The following section begins with a discussion of the broad
  framework of our algorithm DOPING, a moves to an exposition of the
  technical details of the code in \S2. In \S3, we talk about the
  application of DOPING to a test case and corroborate the robustness
  of the algorithm. Application to the real ACS photometry of the
  galaxy vcc9 is also discussed in \S4. \S3 explores the effects of
  varying ambient conditions such as inclination and the assumed
  geometry. \S5 is devoted to discussions and conclusions that are to
  be drawn from the work. In the appendix, a discussion of the
  details of various aspects of DOPING is presented.

  \section{Overview of the Algorithm}
  \label{sec:overview}
  \noindent
  DOPING is a code designed to perform 3-D modelling of systems, given
  2-D images, in a variety of spatial geometries. We iterate over
  trial 3-D luminosity density structures till the best match between
  the projection of the same and the measured 2-D surface brightness is
  obtained. Since deprojection is non-unique unless the intrinsic
  spatial geometry of the system is pinned down, we begin this section
  with a discussion on the motivation and details of how the detailed
  description of the geometry is achieved.

In particular, the true shape and inclination of a system
  can be deciphered, using DOPING, if:
\begin{itemize}
\item the system has a regular geometry, (by which we imply that it
bears an $m$-fold symmetry) and 
\item the relative extent of the system along any three mutually orthogonal
axes are known
\end{itemize}
or if
\begin{itemize}
\item the inclination to the LOS can be varied and the system viewed
at multiple inclinations, in which case 
\item DOPING can perform in {\it any} geometry.
\end{itemize}
Below, we emphasise on deprojection given unknown inclination
  and in the triaxial geometry - the most relevant scenario
  for astrophysical systems. 

In fact, for galaxy clusters, when SZe measurements are available,
  it is possible to measure the extent along all three observer
  coordinate-axes (two axes on the plane of the image and a third
  along the LOS), under the assumption that one photometric axis is
  coincident with a principle axis of the system. Thus, the true
  intrinsic shape and orientation of galaxy clusters can be predicted
  by inverting the X-ray surface brightness map at benchmark deprojection geometries
  \citep{betty_08}. Thus, the luminosity density of galaxy clusters
  can be uniquely determined. An example of this is discussed in \S~ref{sec:A1413}.

However, for triaxial galaxies, the LOS extent is unknown; thus, for
galaxies, the true 3-D shape and orientation cannot be
deciphered. Therefore, for galaxies, we need to assume values for the
polar inclination and the missing axial ratio. The recovery of the 3-D
luminosity density is undertaken, given these assumptions. Also, in
this work, we hold the azimuthal inclination zero.

  It merits mention that our {\it{assumptions are not over-indulgent}}. Any
  deprojection invokes assumptions about the intrinsic geometry and
  the inclinations of the system. Thus, when axisymmetry is assumed, it
  implies that one of the intrinsic axial ratios is held as unity, the
  polar inclination is also assumed and the azimuthal inclination is
  set to zero. This is similar to DOPING in that the user needs to
  assume one inclination and one intrinsic axial ratio for
  galaxies. However, when greater observational information is
  available, as for galaxy clusters, DOPING does not need to invoke
  any assumptions, in contrary to axisymmetry-assuming
  algorithms. 

  The assumptions are designed to be given as inputs for a given run
  of DOPING (Section~2.1 to 2.10). Given that each run of DOPING takes
  about 2 minutes on a 3.2GHz CPU processor , it is possible to scan
  over a wide range of inclination and axial ratio values to record a
  range of corresponding 3-D density distributions. The justification
  of assumptions is discussed in details in \S5.5 and 5.3.

  In order to design a recursive algorithm that can perform
  deprojection in varied geometries, the trial 3-D density structure
  should preferably not be treated as function of a coordinate that
  characterises the geometry at hand. Instead, we need to express the
  3-D density as function of generic coordinates. However, a mapping
  between such generic coordinates and the system geometry is then
  required. This is what we aim for (\S~2.5). In fact, we express this
  mapping by calculating the extent of the system along the LOS,
  through any given point on the image. Such a calculation invokes
  values of all available shape and size related parameters - this is
  discussed in \S~2.5.

  Once this is established, we then discuss
  (\S~2.11) details of how to iterate towards the best
  possible 3-D density structure that projects to the observed surface
  brightness map.

  \subsection{Coordinates used}
  \noindent
  In any kind of deprojection problem, the two coordinate systems that
  suggest themselves readily are the body coordinate frame ($X,Y,Z$)
  and the observer's coordinate frame ($x,y,z$). Here the three
  principal axes of the ellipsoidal system are considered to be along
  the $\hat{\bf X}, \hat{\bf Y}\: \& \:\hat{\bf Z}$ vectors. The LOS
  coordinate is $z$ while the plane of the image is considered scanned
  by the $x-y$ coordinates, i.e the image plane is given by the
  equation $z$=0. The $Z$-axis is considered to be at an inclination
  angle $i$ relative to the line-of-sight, i.e. the $z-$axis.

  The $X,Y,Z$ and $x,y,z$ coordinate systems will be related by two
  consecutive rotations. For triaxial galaxies, {\it neither of these
  rotational angles is an observable in general}. Only when the galaxy
  is highly flattened, can the inclination of its rotational axis to
  the LOS be estimated. The general lack of information about
  inclinations in triaxial systems will need to be compensated for by
  assumptions - while the assumed value of one inclination angle is
  provided as an input to the algorithm, the choice of the other
  inclination angle in this unconstrained situation is chosen to be
  such that our calculations are rendered easy: we assume that one of
  the principle axes lies entirely in the plane of the image. In fact,
  we choose this to be the $X$-axis. Then, the $X$-axis is also a
  photometric axis. We align our observer coordinate system such that
  the $x$-axis lies along the $X$-axis, i.e. $\hat{\bf X} = \hat{\bf
  x}$. Then, $\hat{\bf x}$ is along the photometric major axis for an
  oblate system but along the photometric minor axis in case of a
  prolate system. If the system in hand is triaxial, then there exists
  a scope for a degeneracy, depending on whether the $x$-axis is
  considered the major or minor axis. When the input for the assumed
  value of one inclination is $i$, the equations relating the
  ($X,Y,Z$) system of coordinates to the ($x,y,z$) system are: %
  \begin{eqnarray}
  X &=& x \\ \nonumber
  Y &=& y\cos{i} + z\sin{i} \\ \nonumber
  Z &=& -y\sin{i} + z\cos{i}.
  \label{eqn:transformations}
  \end{eqnarray}
  %

  Thus, {\it {all 3-D density distributions that are recovered by
  DOPING are obtained under the assumption that the azimuthal
  inclination is zero}}. The algorithm can in principle, also work for
  a choice of non-zero azimuthal inclinations; a range of 3-D density
  distributions for a range of choices of this angle is achievable.
  However, in lieu of measured information, the specification of such
  a range is impossible. This is discussed in detail in \S5.3.

  \subsection{Input Data}
  \noindent
  The image or the projection of the galaxy on the plane of the sky is
  treated as built of concentric isophotes. Let the image be built of
  $N_{data}$ number of isophotes and the isophotal annulus between the
  $k-1^{th}$ and $k^{th}$ isophotes be the $k^{th}$ isophotal annulus.
  Here, $k=1,\ldots,N_{data}$.

  The observables that DOPING processes are the characteristics of the
  isophotes, namely, the surface brightness measurement and the
  shape parameters of a given isophote, along with the value of
  its extent along the photometric $x$-axis, i.e. in effect, the surface
  brightness map of the galaxy. Several routines are available for the
  production of such a data set; such as the IRAF implementation of the
  task ELLIPSE \citep{jedrzejewski87}. 

  If the isophotal shape characteristics vary over the extent of the
  image, then it is possible to flag their values according to the
  isophotal annulus that they are observed in; thus, the projected axial
  ratio in the $k^{th}$ isophotal annulus is $q_{pk}$ and the semi-$x$
  axis extent of the $k^{th}$ isophote is $a_{k}$.  Thus, the input data
  table in our work presents: an index for the isophotal annulus or $k$,
  the semi-$x$ axis $a_k$, projected axial ratio $q_{pk}$ and brightness
  values $I_{obs}^k$, for each $k$ i.e. each isophotal annulus that the
  image is binned into. Now, the isophotes of elliptical galaxies are
  often seen to deviate from pure ellipses (e.g. Bender et al. 1988; van
  den Bosch et al. 1994; Ferrarese et al. 2006). Hence, the
  boxiness/diskiness parameters can also be included in the table. To
  keep the introduction of the algorithm simple, in the following
  discussion, we ignore the contribution of these deviations from the
  purely elliptical shape, knowing that these effects can be included
  easily into the isophotal equation as suggested by Jedrzejewski
  (1987). Even more severe deviations from the elliptical isophotal
  shape can be accommodated and these cases (that do not bear a strong
  relevance in astrophysics) are discussed later in \S5.1. The representation
  of isophotes within DOPING is further discussed in AppendixA.

  In the applications of the code discussed in this paper, the position
  angle of the isophotal semi-major axis is assumed constant, though
  scope exists within the algorithm to relax this. The overall scheme
  for such a relaxation is discussed later in \S5.6
  though the incorporation of the same being non-trivial, this will be
  presented within DOPING in a future contribution.

  \subsection{Bayesian Formulation}
  \noindent
  We seek the 3D luminosity density $\rho$ of the system, given the surface
  brightness (surface brightness) maps as the input measurements. The probability of
  spotting the density $\rho$, given the measurements, and all our
  background knowledge (assumptions) about the system ($K$) is:
  \begin{equation}
  \Pr(\rho\vert{SB}, K) \propto \Pr(SB\vert\rho, K)\times{\Pr}(\rho\vert{K}).
  \end{equation}
  This is the Bayesian statement of the problem. The first term on the
  right side of the proportional sign is the likelihood while the second
  term is the prior. 

  In general, the prior that we can use is a uniform one:
  \begin{eqnarray} 
  \Pr(\rho(\cdot)\vert {K}) &=& 1 \quad{\rm {if}\:\rho(\cdot) >= 0} \nonumber \\
		      &=& 0 \quad{\rm {otherwise}},
  \end{eqnarray}
  by which we imply that the prior probability is unity only if
  $\rho(x,y,z)$ is non-negative everywhere.

  However, it is possible that for disk galaxies, $i$ can be established
  from observations. Thus, if the inclination is given as:
  $i=i_0\pm\delta$, then assuming Gaussian errors, our prior will have
  an additional factor that is proportional to $\exp[-(i -i_0)^2/(2\delta^2)]$.

  \subsection{Methodology}
  \noindent 
 A prescribed system geometry would
  imply that the coordinates $x$, $y$ and $z$ are related in terms of
  the shape parameters, i.e. $z$ can be expressed in terms of $x$ and
  $y$. For example, under the assumption of triaxiality, an analytical relation
  links the $x$ and $y$ to the $z$ as $x,y, z$ is a point on the surface of
  a homeoid; this relation takes into account the local values of the
  inclinations and intrinsic axial ratios of the system.

We attempt to express the 3-D density at a point as function of the
  coordinates on the image plane (i.e. $x$ and $y$) and the LOS
  coordinate ($z$) for that point. However, As $x$ and $y$ are
  coordinates on the image plane, they can be measured while $z$ can
  be calculated for a given $x$-$y$ pair, from the aforementioned
  relation between $x$, $y$ and $z$. Once $z$ is established, a trial
  3-D density can be integrated along the LOS, over this established
  range of $z$ values, and the projection compared to the value of surface brightness
  observed at the point $(x, y, 0)$. However, for the aforementioned
  relation to be completely specified, the system geometry needs to be
  invoked. Here we describe how such a relation is specified in the
  astrophysical context.

  For the deprojection of galactic surface brightness maps, if the
  galaxy is considered a single-component system, we assume the galaxy
  to be a {\it{triaxial ellipsoid}}. For multi-component galaxies, such
  as systems with a central component superimposed on an extended disky
  component, {\it{each component is typically ascribed a triaxial
  ellipsoidal geometry}} (Chakrabarty $\&$ McCall, in
  preparation). Isolated off-centred clumps can also be included in the
  modelling (see \S\ref{sec:clumpy}).

  When the galaxy is modelled as triaxial, the details of system
  geometry are described in AppendixB. The crux of the matter is that
  to compensate for our ignorance about the two inclinations and one of
  the two projected axial ratios ($q_p$ and $q_{LOS}$) of an observed
  galaxy, we assume a value for one inclination $i$, set the other
  inclination (angle between ${\bf{X}}$ and ${\bf{x}}$-axis) to zero,
  assume a value for one intrinsic axial ratio $q_1$ and calculate the
  other intrinsic axial ratio $q_2$ from the relation that connects
  $q_2$ to $q_1$, $q_p$ and $i$. The assumptions made to facilitate
  deprojection under triaxiality are similar in number with those made
  when deprojection under axisymmetry is performed (see \S2 and
  \S5.5), although, for deprojection of galaxy clusters, DOPING does
  not need to make such assumptions \citep{betty_08}. Importantly, the
  range of 3-D luminosity density distributions recovered for various
  assumptions, can be gauged with DOPING.

  The axial ratios mentioned above can all vary with distance away
  from system centre. Thus, the assumed axial ratio ($q_1$) can be
  described as any real function of $x$ (as long as the function is
  non-singular over the range covered in the measurements).

  \subsection{Mapping the LOS extent to the system geometry}
  \noindent
  As said before, our aimed deprojection requires evaluation of the
  characteristic extent along the $z$-axis of a point on the image
  plane. To accomplish this, we need to remind ourselves of all the
  relevant image characteristics of the given point on the image
  plane; this includes the surface brightness at this point, the local value of
  projected axial ratio at this point, etc. The determination of the
  $z$-height of a point on the image plane is discussed below and
  graphically represented in Figure~1.

  We put the system on a regular 3-D Cartesian $x-y-z$ grid. We flag
  grid points that lie on the image i.e. the $z$=0 plane, according to the
  isophotal annulus that they lie in. We refer to the $j^{th}$ point inside the
  $k^{th}$ isophotal annulus as $(x_j^k, y_j^k, 0)$. Here
  $j=1,\ldots,N_k$, where $N_k$ is the number of grid
  points with $z$=0, inside the $k^{th}$ isophotal annulus.

  Through the point $(x_j^k, y_j^k, 0)$, let the system extend along the
  $z$-axis, (i.e. the LOS), from $(x_j^k, y_j^k, z_{max}^{jk})$ to
  $(x_j^k, y_j^k, z_{min}^{jk})$. To determine $z_{max}^{jk}$ and
  $z_{min}^{jk}$, we {\it pass a thin triaxial ellipsoidal shell through
  $(x_j^k, y_j^k, z_{max}^{jk})$ and $(x_j^k, y_j^k, z_{min}^{jk})$}. 
  This ellipsoidal shell
  \begin{enumerate}
  \item is centred at (0,0,0),
  \item projects at the assumed inclination $i$, to the $k^{th}$
  elliptical annulus on the image, which in turn has an axial ratio of
  $q_{pk}$ and semi-axis $a_k$ along $\hat{\bf x}$.
  \item has intrinsic axial ratios $q_{1k}$ and $q_{2k}$.
  \item has the points $(x_j^k, y_j^k, z_{max}^{jk})$ and $(x_j^k,
  y_j^k, z_{min}^{jk})$ on it, of which the $x$ and $y$ coordinates are
  known grid points but $z_{max}^{jk}$ and $z_{min}^{jk}$ are
  undetermined.
  \end{enumerate}
  Thus, if we can {\it write the equation of this ellipsoid with $x=x_j^k$,
  $y=y_j^k$ and solve the equation for $z$, the resulting solutions for
  $z$ will be $z_{max}^{jk}$ and $z_{min}^{jk}$}. (Since the equation of
  an ellipsoid is quadratic in $z$, there will be two solutions for $z$).
  To write the equation of the ellipsoid however, we need to know 
  \begin{itemize}
  \item its extent along a principal axis - the extent along the
  $x$-axis, i.e. $X$-axis is known (=$a_{k}$).
  \item the angle between the $Z$-axis and the LOS - $i$ is known by
  assumption.
  \item its intrinsic axial ratios $q_{1k}$ and $q_{2k}$. In the
  absence of measured $q_{LOS}$, $q_{1k}$ is known by assumption.
  $q_{2k}$ is derived as follows.
  \end{itemize}
  $q_{pk}$ is related to $q_{1k}$ and $q_{2k}$ through a simple
  geometrical relation that also involves $i$:
  \begin{equation}
  \displaystyle{q_{pk}^2} = \displaystyle{\frac{{q_{1k}^2}q_{2k}^2}
					    {q_{1k}^2\cos^2(i) + q_{2k}^2\sin^2(i)}
				      }.
  \label{eqn:ep}
  \end{equation}
  This relation gives $q_2^{k}$ \citep[see][]{betty_08}. In this way, we
  constrain all parameters that define the ellipsoidal shell that passes
  through the points $(x_j^k, y_j^k, z_{max}^{jk})$ and $(x_j^k, y_j^k,
  z_{min}^{jk})$ (points I to IV of first list in this subsection). The
  equation of such a fully characterised ellipsoid is
  \begin{eqnarray}
  \displaystyle{
		(x_k^j)^2 +(y_k^j)^2\displaystyle{({q_{2k}^2}{\cos{^2}{i}} + {q_{1k}^2}{\sin{^2}{i}})}+}
  & &  {} \nonumber \\	      
  \displaystyle{
		z^2\displaystyle{({q_{2k}^2}{\sin{^2}{i}} + {q_{1k}^2}{\cos{^2}{i}})}
  -y_k^jz\sin{2i}\displaystyle{({q_{1k}^2} - {q_{2k}^2}) }}
   &=& a_k^2. 
  \label{eqn:ell_i}
  \end{eqnarray}
  Here $q_1^{k}, q_2^{k}, a_k$ and $i$ are all known. $x_k^j$ and
  $y_k^j$ are known since the point $x_k^j, y_k^j, 0$ has been
  identified to lie inside the $k^{th}$ isophotal annulus. Thus, this
  quadratic equation is solved for $z$ and its two solutions are
  $z_{max}^{jk}$ and $z_{min}^{jk}$.

\begin{figure}
  \includegraphics[width=8cm, angle=0]{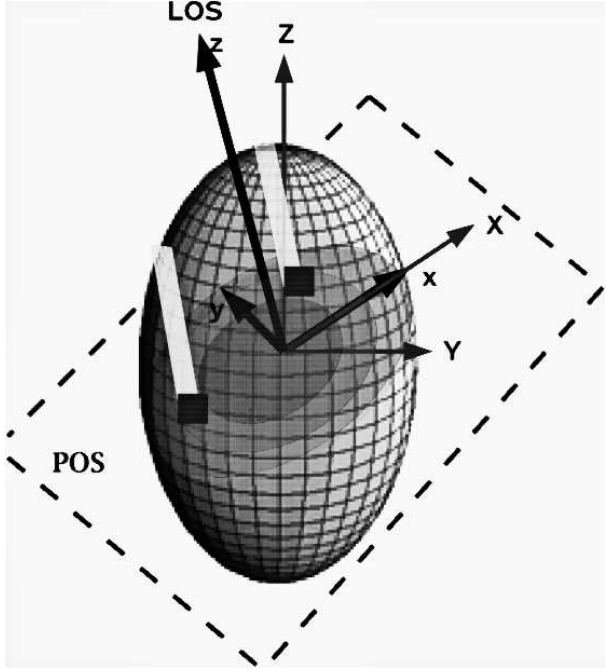}
  \caption{Figure showing geometrical considerations adopted in the
    design of the algorithm. The system is represented as the
    ellipsoid. The $X$, $Y$ and $Z$ axes (in thin black lines)
    represent the three principle axes of the system while $x$ and $y$
    mark the photometric axes and the $z$-axis is the LOS (in thicker
    black lines). A rectangular section of the image plane (i.e. the
    $z$=0 plane) is represented by the tilted rectangle in the broken
    lines; this plane cuts the ellipsoid in an elliptical disk which
    is depicted by the translucent gray disk. Generic isophotal
    annuli on this disk are depicted in centrally increasing
    gray-scale intensity. Two generic points, lying inside the
    intermediate isophotal annulus, are shown as the two black
    squares. The extent of the system along the positive $z$-axis, at
    these two marked points are represented by the lengths of the
    white rectangles that are oriented parallel to the LOS. In the
    text, one such point, generically considered to be inside the
    $k^{th}$ isophotal annulus, is referred to as $(x_j^k, y_j^k, 0)$,
    while the tip of the white rectangle emanating from this point is
    ascribed coordinates $(x_j^k, y_j^k, z_{max}^{jk})$.  The extent
    along the negative $z$-axis is not shown in this diagram. The
    ellipsoidal shell that passes through $(x_j^k, y_j^k,
    z_{max}^{jk})$ is fully constrained (see \S~2.5). Using $x_j^k$
    and $y_j^k$, (coordinates of the known point on the image plane),
    the intrinsic shape parameters of the $k^{th}$ isophotal annulus
    ($a_k$, $q_{1k}$, $q_{2k}$) and the inclination $i$, the equation
    of this ellipsoid is solved to obtain the value of $z_{max}^{jk}$
    (and $z_{min}^{jk}$). Thus, the $z$-extent through which a trial
    3-D $\rho(x_j^k, y_j^k, z)$ has to be projected, is ascertained,
    for all $j$ and $k$.
  \label{fig:scheme}}
  \end{figure}

  In this way, the extent of the system, along the $z$-axis, through any
  point on the 2-D image is determined. A schematic of this procedure is
  presented in Figure~\ref{fig:scheme}. We now continue with the nomenclature
  introduced in this section, to describe generic points lying inside generic
  isophotal annuli.

  \subsection{$z-histogram$s}
  \label{sec:zhisto}
  \noindent
  In order to keep the formalism {\it flexible}, we seek a form of the 
  3-D density in terms of $z$.

  Thus, we discretise the density structure $\rho(x_j^k, y_j^k, z)$
  where $z\in[z_{min}^{jk}, z_{max}^{jk}]$, $\forall\: j, k$.  Actually,
  in our work, we bin the range $[0, z_{max}^{jk}]$, and find the
  luminosity density $\rho(\cdot)$ for only one half of the system (at
  positive $z$ only). The density for the other half is given using
  the symmetry argument $\rho(x,y,z)=\rho(-x,-y,-z)$, which is valid
  under the assumption of triaxial geometry. In other words, we invert
  the projection integral $I(x_k^j, y_k^j) =
  2\int_{0}^{z_{max}}\rho(x_k^j, y_k^j, z) dz$ instead of $I(x_k^j,
  y_k^j) = \int_{z_{min}}^{z_{max}}\rho(x_k^j, y_k^j, z) dz$.

  We do this by binning the $z$-range between $z_{min}^{jk}$ and
  $z_{max}^{jk}$. The binning is {\it {logarithmic}} since the
  measurements of surface brightness values are typically obtained for
  an astronomical system at increasingly wider isophotal
  annuli. $\rho(x_j^k, y_j^k, z)$ is held a constant over each
  $z$-bin. Thus, the density structure along the $z$-axis, through the
  point $(x_j^k, y_j^k, 0)$ on the image, looks like a 1-D histogram. We refer to
  this construction as the ${\bf{z-histogram}}$, corresponding to the
  point $(x_j^k, y_j^k, 0)$. The $z$-range of a ${\bf{z-histogram}}$ spans
  the interval: $z_{min}^{jk}$ and $z_{max}^{jk}$. Thus, this $z$-range
  depends on the point on the image through which the ${\bf{z-histogram}}$ is
  constructed.

  \subsection{Why $z-histogram$s instead of $\xi-histograms$s}
  \label{sec:xihisto}
  \noindent
  We choose the basis of $\rho$ to be $z$ instead of a function of the
  system shape such as the ellipsoidal radius $\xi$. Reliance of the
  deprojection of the observed surface brightness map on the intrinsic shape would
  curb the reach of the algorithm in the following two ways:
  \begin{itemize}
  \item systems with different geometries that cannot be ascribed a
    general triaxial shape cannot then be tracked by the same code. An
    example of such a system within the astrophysical context could be a
    galaxy that is better described by a cylindrical intrinsic shape,
    such as the LMC. surface brightness of such a galaxy can however be deprojected
    under the flexible DOPING, with minimal changes imposed on the
    algorithm. In this non-triaxial geometry, the calculation of the
    intrinsic axial ratios from the measured projected axial ratios (and
    hence the calculation of the $z$-height of any point on the image) is
    different from the triaxial case; these calculations are
    performed within a modular sub-routine, before the iterations begin.
    The rest of the algorithm (iterative search for the most likely
    density structure) is unaffected by the difference in the system
    geometry. Thus, the same code can be used to undertake deprojection
    in general geometries.
  \item luminosity density distributions of systems with imposed
    substructure or extra galactic components that are imposed on the
    background galactic structure (such as disk+bulge systems) cannot be
    obtained in a single-step, integrated fashion if the algorithm is
    designed exclusively within the geometry of the background structure.
  \end{itemize}
  However, it may be argued that the choice of $z$ as the basis function
  for $\rho$ will subvert the underlying geometry under which we
  deproject and the independent updating of the ${\bf{z-histograms}}$
  will render the resulting density structure noisy. This is not the
  case. The system geometry is reinforced via:
  \begin{itemize}
  \item the determination of the $z$-height of a given point on the
  image plane. This value robustly reflects the intrinsic geometry of
  deprojection.
  \item penalising all solutions for $\rho$ that do not adhere to a form
    in which there is {\it maximum variance between densities that are at
    different ellipsoidal radii but minimum variance between densities
    at the same $\xi$}. This characteristic can be incorporated by
    introducing a penalty function that is proportional to the Laplacian
    of the current choice of $\rho$, where the Laplacian operator
    involves differentiation w.r.t. $\xi$. This is discussed in detail in
    the following subsection.
  \end{itemize}

  \subsection{Laplacian regularisation}
  \label{sec:laplacian}
  \noindent
  We understand that the sought solution for $\rho(x,y,z)$, as given
  by the assembly of $z$-histograms, is in need of regularisation. We
  choose to introduce this regularisation such that we achieve
  low-dimensional representation of higher-dimensional information. In
  particular, we are interested to recover density that is a function
  of $\xi$, i.e. we work with a penalty function that reflects the
  intrinsic geometric structure of the input space \citep{haykin_08,
  wang_06}\footnotemark.

  \footnotetext{Such motivation is also found in {\it Maximum Variance Unfolding} that also implements Laplacian regularisation and has been studied by \cite{weinberger_06, sha_05, sun_06}. }

  Such sought, {\it similarity based smoothing} is ensured by adopting a
  penalty function ${\cal{P}}$ that is given in terms of the Laplacian 
  of the object function:
  \begin{eqnarray}
  {\cal{P}} &=& \alpha\displaystyle{(\nabla^2 \rho)^2} \quad {\textrm{where}} \nonumber \\
  \nabla^2 &\equiv& \displaystyle{\frac{\partial^2}{\partial^2\xi}}
  \label{eqn:xi}
  \end{eqnarray}
  Here $\alpha > 0$ is a regularisation parameter, that reflects a 
  trade-off between maximising variance between $\rho$ at different $\xi$
  and reinforcing the relevant deprojection geometry.

  \subsection{Density structure}
  \label{sec:zhisto2}
  \noindent
  DOPING works recursively, via an inverse approach. At every iterative
  step, a trial $\bf{z-histogram}$ is chosen for each grid point on the
  image, i.e. for a given $j$, $k$. Each such $\bf{z-histogram}$ is
  updated independently during an iterative step, to render the whole
  3-D density structure of the galaxy updated. Such updating of the
  $\bf{z-histogram}$ is done while maintaining positivity of $\rho$.

  Once updated, the density structure is projected on the $z$=0 plane
  and this projection is compared to the surface brightness data. This comparison
  defines the likelihood function which is maximised for the best
  match. The likelihood is supplemented with a penalty that was
  discussed in the last paragraph. The global maxima of the
  likelihood function is sought by our MCMC algorithm to yield the
  most likely density structure, given the surface brightness data. However, we
  choose only those solutions which are ``smooth'', as dictated by the
  used regularisation scheme.

  \subsection{Likelihood}
  \noindent
  The probability of the data given the model - i.e. the observed
  surface brightness map, given a trial 3-D density - is expected to
  be normal. This is reinforced on the basis of the following.

  The likelihood or the probability of a measured surface
  brightness map, given a choice of the 3-D density structure, has to
  be a function of the distance between the projection of the 3-D
  density on the image plane and the measured surface brightness. In
  particular, $\Pr(data|\rho(\cdot))$, is such that when the
  projection of $\rho(\cdot)$ on the image plane is concurrent with
  the measured surface brightness distribution,
  $\Pr(data|\rho(\cdot))$=1. Additionally, the further is
  $\int\rho(x_k^j, y_k^j, z) dz$ from the surface brightness
  measurement $I_{obs}^k$ in the $k^{th}$ isophotal annulus, the
  smaller is the likelihood; in fact, for $\vert \int\rho(x_k^j,
  y_k^j, z) dz - I_{obs}^k \vert \longrightarrow \infty$,
  $\Pr(data|\rho(\cdot))$=0. Since the likelihood is a function of the
  absolute distance between $\int\rho(x_k^j, y_k^j, z) dz$ and
  $I_{obs}^k$, (for any $k$), for two different 3-D densities
  $\rho_1(x_K^j, y_k^j, z)$ and $\rho_2(x_K^j, y_k^j, z)$, if $\vert
  \int\rho_1(x_k^j, y_k^j, z) dz - I_{obs}^k \vert = \vert
  \int\rho_2(x_k^j, y_k^j, z) dz - I_{obs}^k \vert $, it implies that
  $\Pr(I_{obs}^k|\rho_1(\cdot)) = \Pr(I_{obs}^k|\rho_2(\cdot))$,
  i.e. the likelihood is symmetric about $I_{obs}^k \: \forall \:k$.
  Also, for two different surface brightness measurements,
  $I1_{obs}^k$ and $I2_{obs}^k$, if the likelihood corresponds to the
  same value of $\int\rho(x_k^j, y_k^j, z) dz$, it implies that
  $I1_{obs}^k=I2_{obs}^k, \: \forall k$. Given these to be the only
  constraints on our choice of the likelihood ${\cal{L}}$, it is
  sufficient to consider the distribution $\Pr(I_{obs}^k|\rho(\cdot))$
  to be normal - proportional to a Gaussian of the form
  $\exp[-(I_{obs}^k - \int\rho(x_k^j, y_k^j, z) dz)^2/(I_{obs}^k)^2]$,
  where the denominator in the exponential is a scale that is invoked
  to ensure a dimensionless term; the measurement offers a ready
  scale. In, details, the log likelihood is
  \begin{equation}
  \ln{\cal{L}} = \displaystyle{-\sum_{k=1}^{N_{data}}{\frac{1}{N_k}{\sum_{j=1}^{N_k}{\frac{(I_k^{j} - I_{obs}^k)^2}{(I_{obs}^k)^2}}}} - {\cal{P'}}},
  \label{eqn:nabla}
  \end{equation}
  where $N_k$ is the number of ${\bf{z-histograms}}$ for the $k^{th}$
  isophotal annulus (see \S\ref{sec:nk}). ${\cal{P'}}$
  represents the penalty function, defined along the lines of ${\cal{P}}$ of Equation~\ref{eqn:xi}:
  \begin{eqnarray}
  {\cal{P'}} &=& \displaystyle{\sum_{k=1}^{N_{data}}{{\sum_{j=1}^{N_k}{\alpha\nabla_{jk}^2\rho(x_k^j, y_k^j, z_k^j)^2}}}}, \quad {\textrm{where}} \nonumber \\
  \nabla_{jk}^2 &\equiv& \displaystyle {\frac{\partial^2}{\partial\xi_{jk}^2}} 
  \end{eqnarray} 
  Here ($x_k^j, y_k^j, z_k^j$) is a point at which a value of the density
  is defined and $\xi_{jk}^2$ is given by the left-hand-side of
  Equation~\ref{eqn:ell_i}, with $z$ replaced by $z_k^j$.

  \subsection{Interval Estimation of 3-D Density}
  \label{sec:errors}
  \noindent
    We choose to implement MCMC optimisation with the
    Metropolis-Hastings algorithm \citep{hastings, metropolis,
    tierney, tanner, gelman_book}. The set of models
    identified by our optimiser in the maximal region of the
    likelihood is really an ensemble of {\it all the ${\bf
    z-histograms}$ corresponding to each of the grid points on the
    image plane}, i.e. the full 3-D density structure. (see Appendix~D
    for greater details of the optimisation procedure and the choice
    of the MCMC parameters).  Thus, the dimensionality of the
    likelihood function is the product of the number of bins along
    each of three spatial axes. When the algorithm identifies the
    maximal region of the likelihood function, ${\bf z-histograms}$
    corresponding to this maximal region are recorded. The 3-D density
    distributions given by this set of ${\bf z-histograms}$ are
    (identified and for a given $(x,y,z)$, the values of $\rho(x,y,z)$
    from these identified density distributions) are sorted. The
    $\pm$1-$\sigma$ range of values of density, about the medial
    density at this point is recorded. Such a range of values of
    density, over all $x$, $y$ and $z$ then defines the most likely
    3-D density structure that we identify as corresponding to the
    surface brightness data at hand. The implementational details of our interval
    estimation of luminosity density at a given point $(x,y,z)$ is
    discussed in Appendix~D.1, D.2 and D.3.

  \subsection{Construction of Seed or Trial Luminosity Density}
  \label{sec:nk}
  \noindent
  In the very first iterative step, the density is ascribed a
  arbitrary functional form $\rho_{seed}(x, y, z)$ - the final answer
  should be independent of this choice of the initial guess for the
  density or the seed density. We use crude estimates of the
  parameters that define $\rho_{seed}$ to begin multiple runs.

  \section{Testing $\&$ Applications}
  \label{sec:results}
  \noindent
  In this section, we present the results of our analysis done with
  simulated data sets that have been designed to mimic the brightness
  distributions of disk-like and elliptical galaxies with rapidly
  varying eccentricity profiles, that achieve very high eccentricities
  indeed. Our examples include
  \begin{itemize}
  \item Test~I: a system that resembles a razor-thin disc with a small
    (of scale length of 0\Sec5 as compared to he extent of the system
    which is about 100$^{''}$) round component resembling a tiny bulge
    embedded in the centre. The eccentricity evolves from zero at the
    centre to about 0.95 by 2\Sec0 and by 3\Sec0, is then maintained
    at nearly unity. The radial run of the eccentricity of this system
    is represented in filled circles in Figure~\ref{fig:ecc}. Thus,
    this system, if tested favourably with DOPING, will validate the
    following characteristics of the algorithm:
  \begin{itemize}
  \item is able to deal with galaxies of varying morphologies, including
  disk galaxies.
  \item is robust even when eccentricity is as high as nearly unity.
  \item is able to deal with very rapid rise in eccentricity.
  \end{itemize}

  \item Test~II: a system that is rounder in the centre but the
  eccentricity of which rises to about 0.97, over a length scale of
  40$^{''}$. Thus, this is an elliptical galaxy with widely varying
  intrinsic shape; the axial ratio changes from nearly 0 at the centre
  to about 7 at the outer edge of the system. The radial eccentricity
  profile of this galaxy with widely varying intrinsic shape, is shown
  in open circles in Figure~\ref{fig:ecc}. This example reinforces
  DOPING's efficacy in describing systems with different morphologies.
  \end{itemize}

  The brightness maps of these test cases will be input into DOPING.
  Such toy surface brightness are constructed as the LOS projection of chosen
  analytical luminosity density distributions
  (Equation~\ref{eqn:density}, see below). The projection is performed
  for a given analytical choice of the eccentricity $e$. The luminosity
  density distributions that DOPING recovers are then compared to the
  known analytical density from which the test surface brightness maps were
  extracted. 

  The deprojection in this section is performed under the assumptions of
  oblateness and edge-on viewing, i.e. $i=90^{\circ}$. Therefore, for the
  test galaxies, $q_1^k=1\!\forall\:\! k=1,\ldots,N_{data}$ and $q_{pk} =
  q_{2}^k\!\forall\:\! k$, i.e. the projected and intrinsic eccentricities
  concur.

  \subsection{Recovery of a Known Density Distribution}
  \label{sec:test}
\begin{figure}
  \centering{
  \includegraphics[width=7cm]{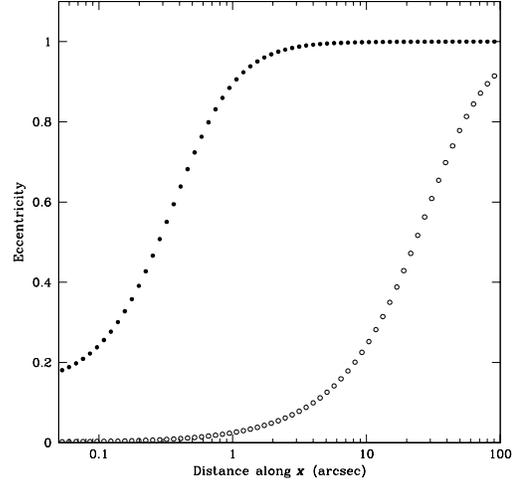}}
  \caption{The chosen eccentricity profile of the test galaxy Test~I,
  shown as a function of $r$, in filled circles. The same for Test~II is
  shown in open circles.
  \label{fig:ecc}}
  \end{figure}

  \noindent
  The surface brightness\footnotemark and projected eccentricity
  profiles which constitute our test data sets are discussed here. The
  intrinsic eccentricity is as shown in Figure~\ref{fig:ecc}. The run of
  eccentricity with radius $r (= \sqrt{x^2+y^2+z^2})$ is given as
  follows:
  \begin{equation}
  e(r) = \displaystyle{\sqrt{1 - \displaystyle{\frac{1}{1+\displaystyle{\frac{r^2}{r_c^2}}}}}},
  \label{eqn:ecc}
  \end{equation}
  where $r_c$ is a scale length. The eccentricity is chosen to be
  function of the spherical radius $r$, rather than the major axis
  coordinate, in order to ease the calculation of the projection
  integral leading to the formulation of the test surface brightness.
  The analytical luminosity profile, from which the brightness
  data has been extracted, is
  \begin{equation}
  \rho(x, y, z) = \displaystyle{\frac{A}{\left(B^2 + x^2 + \displaystyle{\frac{y^2}{1-e(r)^2}} +z^2\right)^{1.5}}},
  \label{eqn:density}
  \end{equation}
  with $e(r)$ given in equation Eqn~\ref{eqn:ecc}. $B$ is a scale length
  or core radius and $A$ is the central density scaled by the factor
  $B^{3}$.

  \footnotetext{Note that although we will use units of magnitude for
    the surface brightness and density profile, DOPING works in linear
    units of intensity.}

  We integrate $\rho(x, y, z)$ along $z$, after plugging in the form of
  $e(r)$ from Equation~\ref{eqn:ecc}, into Equation~\ref{eqn:density}.
  The result of this integration is the toy surface brightness data that
  we want DOPING to invert.
  \begin{equation}
  I (x, y) = 2\displaystyle{\frac{A}{\left[B^2 + 
					  (x^2+y^2)
					  \left(1+
						\displaystyle{\frac{y^2}{r_c^2}}
					  \right)\right]
					  \sqrt{1+
						\displaystyle{\frac{y^2}{r_c^2}}}
					  }}
  \label{eqn:I}
  \end{equation}
  In the toy data set Test~I, the surface brightness profile is sampled
  at 64 locations along the galaxy semi-major axis, from 0.\sec05 to
  about 100\sec, while for Test~II, the binning is about 3 times
  finer. This is done to demonstrate that our code can be applied to a
  data set of any length without any additional modifications. The
  surface brightness forms the third column of the input data table (the
  first two columns are the semi-major axis and eccentricity). Two
  additional columns contain measurement errors in the brightness and
  eccentricity; it is possible to incorporate the measurement errors,
  assuming a Gaussian error distribution. However, such errors are
  typically negligible compared to the errors in the profiles recovered
  by DOPING (\S~\ref{sec:errors}).
  \begin{figure*}
  {\begin{center}
  \includegraphics[scale=.85]{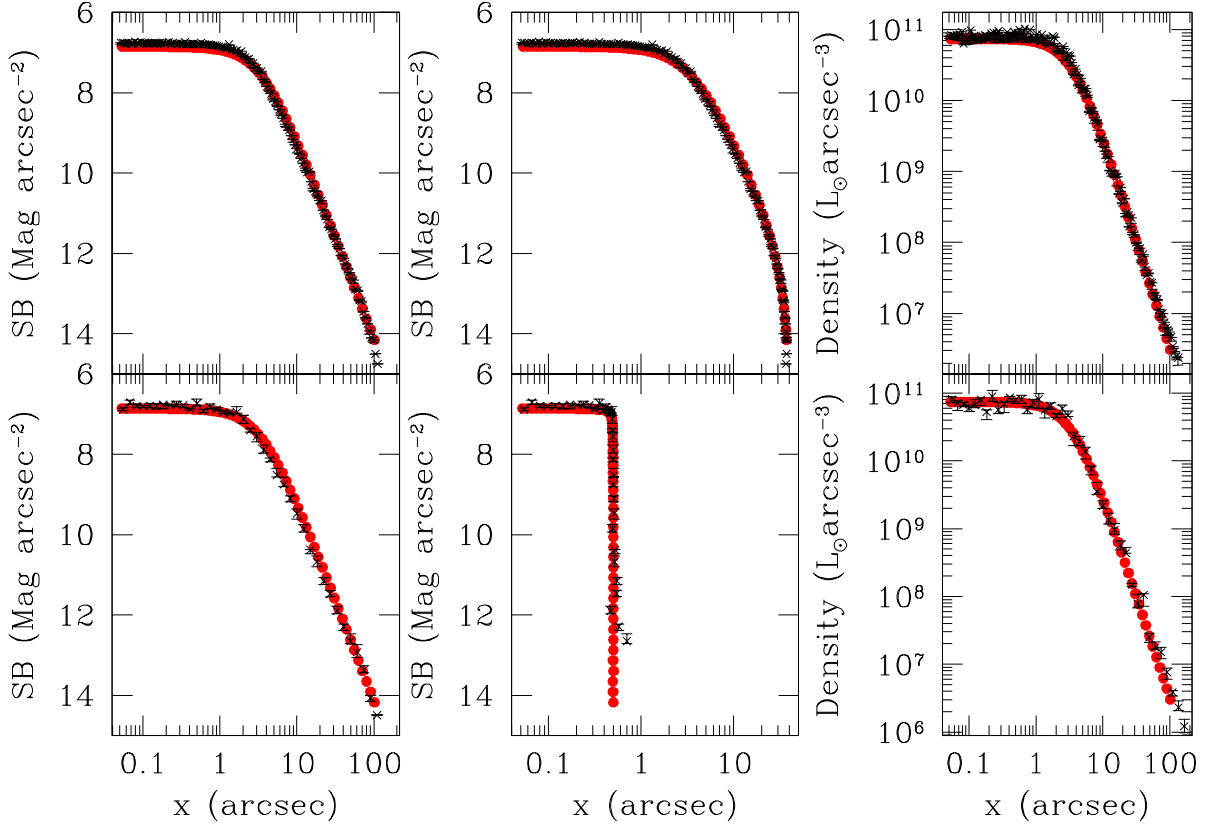}
  \end{center}}
  \caption{Figure to display the performance of DOPING in the simulated
  test cases Test~I (a disk galaxy with a round ellipsoidal centre that
  extends to only about 0.\sec5) in the lower panels and Test~II (an
  elliptical galaxy with varying eccentricity) in the upper panels. The
  chosen analytical density distributions is shown in red open circles,
  as a function of the major axis coordinate, in the right panels. This
  known density, along with the eccentricity profile shown in
  Figure~\ref{fig:ecc}, implies the surface brightness distributions
  which are shown in red along different azimuths $\phi$, (major axis,
  minor axis), in red filled circles, in the left and middle panels. The
  recovered density distributions along the major axis are over-plotted
  on the known density profiles, in black, in the right panels.
  Projections of these recovered density distributions are shown in
  black, as functions of $x$ along two different azimuths ($\phi=0$,
  i.e. major axis and $\phi=90^{\circ}$, i.e. minor axis), superimposed
  on the brightness data along these respective azimuths. The sharp rise
  of the eccentricity profiles in the two test cases (over length scales
  that are a factor of 80 apart) is indicated by the much sharper
  decline of the density along the minor axis, compared to the major.
  Higher regularisation is implemented in the recovery of the density
  distribution in the Test~II case than in the case of Test~I. In this
  figure, as well as in all subsequent figures, we assume that the
  surface brightness is measured in the SDSS $z$-band for a galaxy at a
  distance of 17Mpc. This is true unless otherwise stated.}
  \label{fig:testcase}
  \end{figure*}

\begin{figure*}
  {\begin{center}
  \includegraphics[scale=.9]{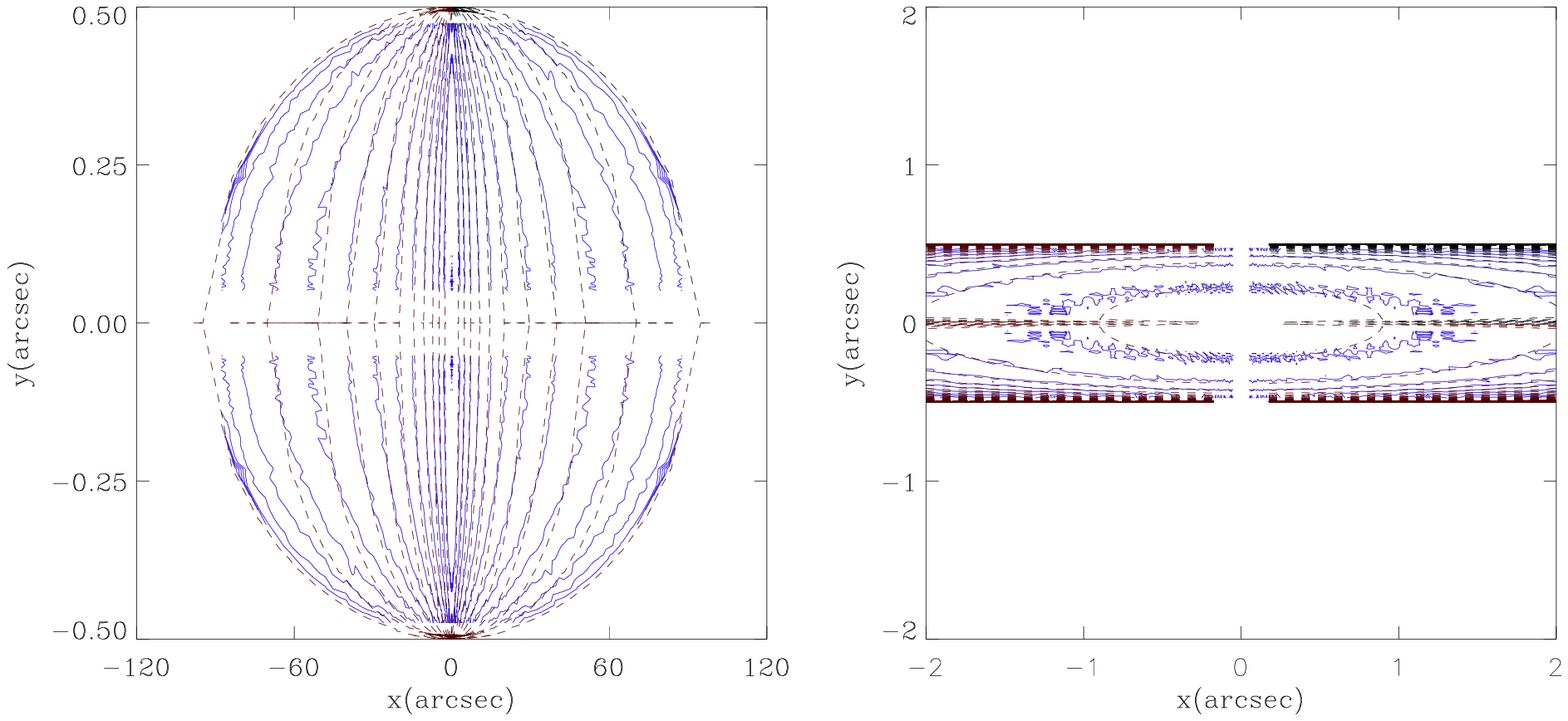}
  \end{center}}
  \vskip1.cm
  \caption{Left: the 2-D surface brightness (in mag/arcsec$^2$)
  distribution of our flat test galaxy Test~I, as a contour plot on the
  plane of the sky ($x-y$ plane). The contours in broken lines pertain
  to the toy brightness data that was fed into DOPING while the solid
  lines represent the projection of the 3-D luminosity density that
  DOPING recovers. The gap around $y$=0 occurs in the distribution of
  the projected density since the smallest (logarithmic) spatial bin is
  about 1pixel, i.e.  0$^{''}$.05. Right: same as for the left panel,
  except that in this case, the central rounder part of the test galaxy
  has been focused upon.}
  \label{fig:contour_initial}
  \end{figure*}

\begin{figure}
  \centering{
  \includegraphics[scale=0.45]{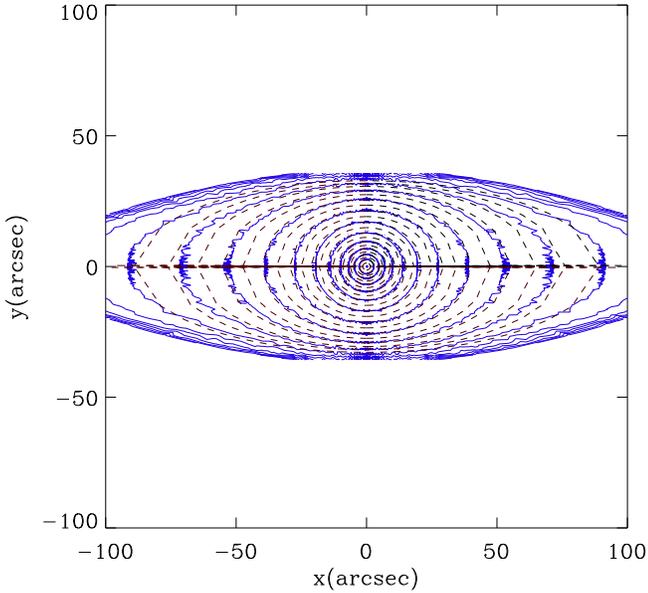}}
  \caption{2-D surface brightness (in mag/arcsec$^2$) distribution of
  the elliptical test galaxy Test~II, compared to the plane of the sky
  projection of the luminosity distribution recovered by DOPING for this
  system. 
  \label{fig:contour_rc100}}
  \end{figure}

  The robustness of the comparison between the test 2-D brightness
  distribution of the test galaxies and the projections of the recovered
  density distributions is brought out in
  Figure~\ref{fig:contour_initial} and Figure~\ref{fig:contour_rc100}.

  The recovered density profile as well as its projection appear to
  tally very favourably with the known distributions.

  \subsection{Changing Inclinations}
  \label{sec:inclination}
  \noindent
  In this section, we investigate the extent to which the recovered
  luminosity density is rendered uncertain by our ignorance of the
  inclination angle $i$, under a given assumption about the geometry
  of the system and for a given set of observables. 

  In order to track this uncertainty, we use the test surface brightness data given in
  Equation~\ref{eqn:I}, and constrain the projected eccentricity to be
  radially invariant: $e_p = e_{p0}$. Working with a constant
  $e_p$ is preferred to a radially dependent projected eccentricity, on
  grounds of ease of interpretation of the results. The deprojection of
  the test galaxy is performed under the assumption that the galaxy is
  oblate in shape. For such a geometry, the inclination cannot be less
  than $\sin^{-1}e_{p0}$.

  We perform a suite of deprojections of the test surface brightness data with $i$
  set to $i_1,i_2,i_3,\cdots,90$\deg, where $i_1$ is the minimum
  inclination consistent with the observed projected eccentricity of
  $e_{p0}$, i.e.  $i_1=\sin^{-1}e_{p0}$. Here $e_{p0}$ is chosen to be
  one of the following 4 values: $e_{p0} = 0, 0.71, 0.87, 0.95$. These
  values of $e_{p0}$ were chosen to span the range that early type
  galaxies are typically observed to bear. Deprojections were
  performed for each $e_{p0}$, at each of the 4 selected $i$. Thus,
  our experiments can track 4 test galaxies which are distinct in
  their flattening, each assumed oblate and viewed at a suite of
  different inclinations, the smallest of which is set by the
  projected eccentricity.

\begin{figure}
  \centering{
  \includegraphics[scale=.4]{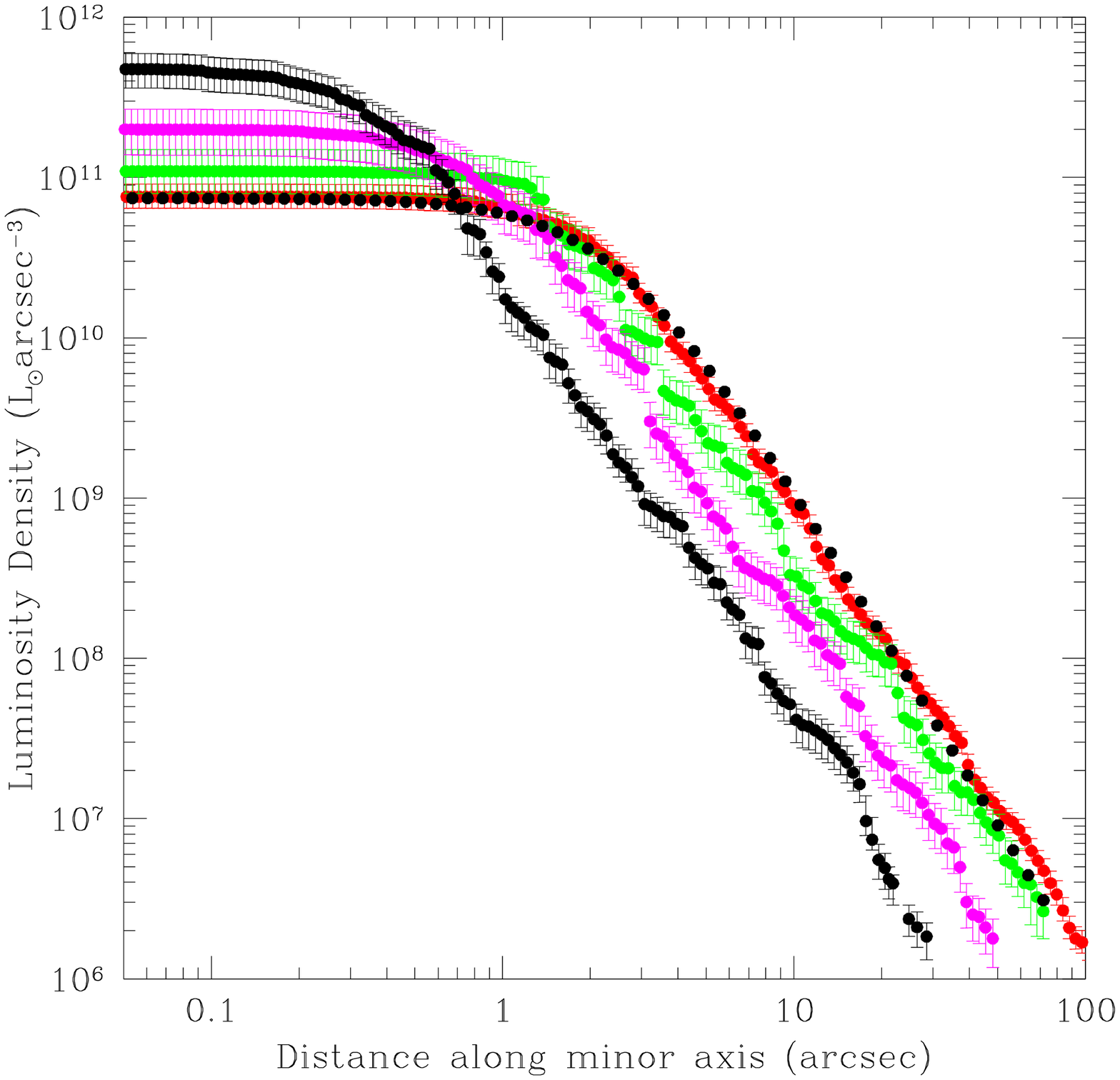}}
  \caption{Figure showing luminosity density distributions recovered
    by deprojecting the surface brightness profile given by
    Equation~\ref{eqn:I}, under the assumption of oblateness, given a
    projected eccentricity of 0.71, viewed at inclinations of about
    46\deg~(in black), 48\deg~(in magenta), 55\deg~(in green) and
    90\deg~(in red). The recovered density distributions are plotted
    along the minor axes. The true luminosity density of the test
    galaxy is shown in filled black circles. The profiles above were
    recovered from runs performed with various bin widths ad smoothing
    parameter values. The estimation of the error bars on the
    recovered density profiles is discussed in Appendix D.1.
  \label{fig:ep71}}
  \end{figure}

  Figure~\ref{fig:ep71} shows the density profiles recovered by DOPING
  by deprojecting the test surface brightness (Equation~\ref{eqn:I}), for the choice
  of $e_{p0}$=0.71. This corresponds to $i_1\approx 45$\deg. For this
  configuration, deprojection is performed at four distinct values of
  the viewing angle, in the range of [45\deg, 90\deg], at $i\approx
  46\deg,~48\deg,~55\deg,~90\deg$.  It is possible to obtain the given
  surface brightness map that manifests a given projected flatness ($e_{p0}$=0.71) at
  these 4 different inclinations, only by projecting the luminosity
  densities of 4 distinct oblate galaxies with intrinsic eccentricities
  of 0.99, 0.95, 0.87, 0.71. 

  Thus, deprojection of the observed brightness map, carried out at
  varying inclinations, is characterised by variation in amplitudes as
  well as shapes. However, it is only along the major axes (the
  semi-axis along {\bf {$\hat{x}$}}) that the deprojected profiles will
  appear similar in shape but different in amplitude. This owes to our
  definition of the toy surface brightness distribution
  (Equation~\ref{eqn:I}). Along all other directions, the recovered
  density distributions will manifest differences in shape as well. It
  is for this reason that in Figure~\ref{fig:ep71} we present the
  recovered density profiles along the galaxy minor axes. The variation
  in shape across the set of deprojected density profiles is clear from
  this figure.

  It is to be noted that the projections of the recovered density
  profiles coincide with the input surface brightness data in each case. However, once
  $i\leq58$\deg~ (for projected eccentricity =0.71), the 3-D density
  profiles become a sensitive function of $i$.  As expected, the
  recovered density is maximum (at every radius) when the intrinsic
  eccentricity is highest (i.e., the inclination angle is lowest). When
  the galaxy is assigned an even higher projected eccentricity, the
  uncertainty in the obtained density shows up at even lower $i$, i.e at
  inclinations closer to the face-on configuration.

\begin{figure}
  \centering{
  \includegraphics[scale=0.4]{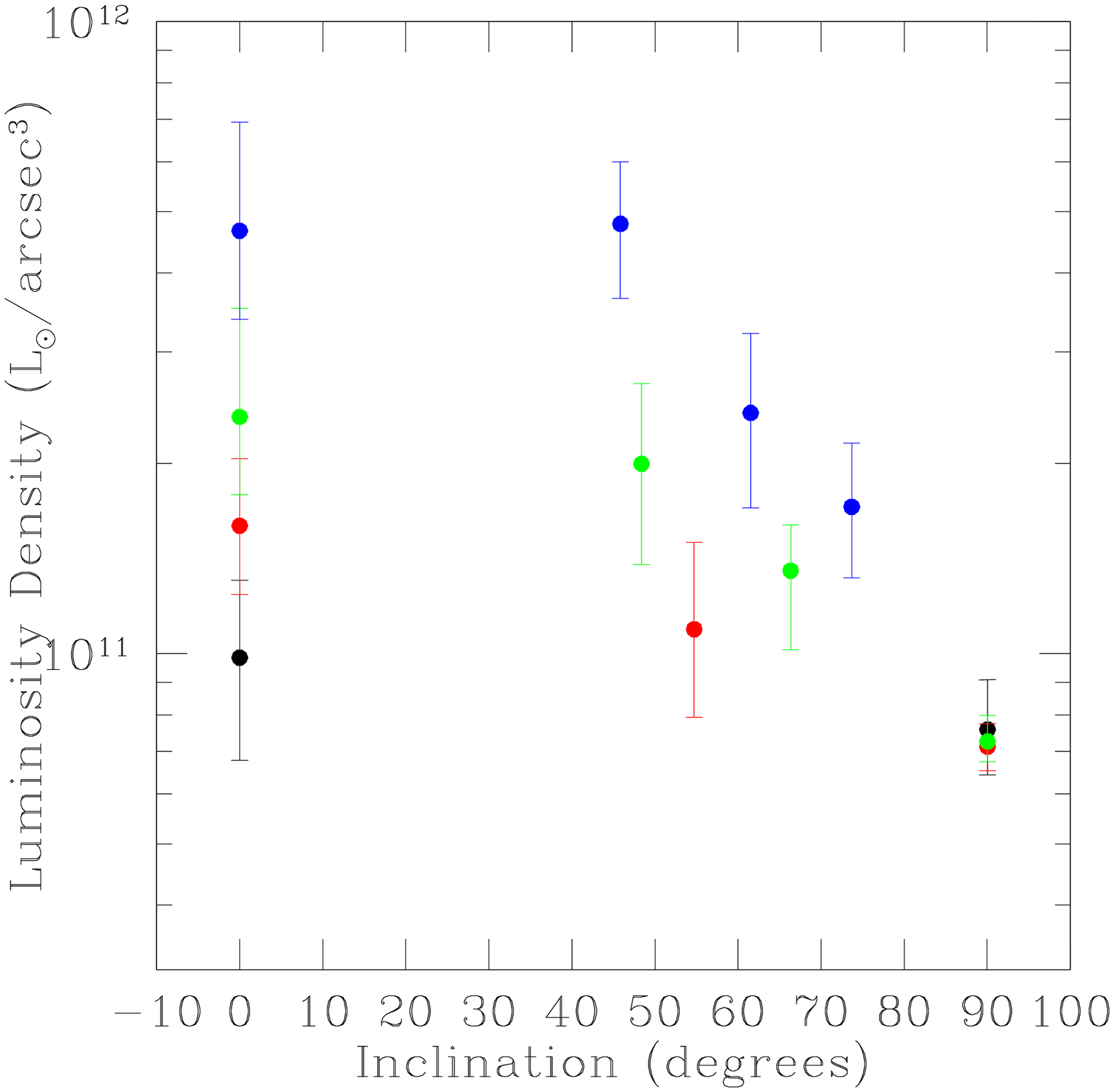}}
  \caption{Central luminosity density, plotted as a function of
  inclination for four different values of the intrinsic eccentricity.
  When the intrinsic eccentricity is 0.71, the obtained central density
  points are shown in black. The colour coding for the other values of
  $e$ is as follows: $e$=0.87, 0.95 and 0.99 correspond to red, green
  and blue, respectively. The case of inclination=0 obviously indicates
  the situation when the observed isophotes are circular, i.e. the
  observed projected eccentricity is zero.
  \label{fig:cenrho}}
  \end{figure}

  Figure~\ref{fig:cenrho} presents the value of the recovered luminosity
  density at the innermost radial bin (about 0.\sec05), plotted as a
  function of the assumed inclination, for varying values of the
  intrinsic eccentricity, under the assumption of oblateness. As
  expected, the central density values concur (within the error bars),
  for the edge-on configuration, while density is highest at the centre
  at $i=0$\deg~, for the intrinsically most eccentric system.

  \subsection{Changing Geometry}
  \label{sec:sphericity}
\begin{figure}
  \centering{
  \includegraphics[scale=0.4]{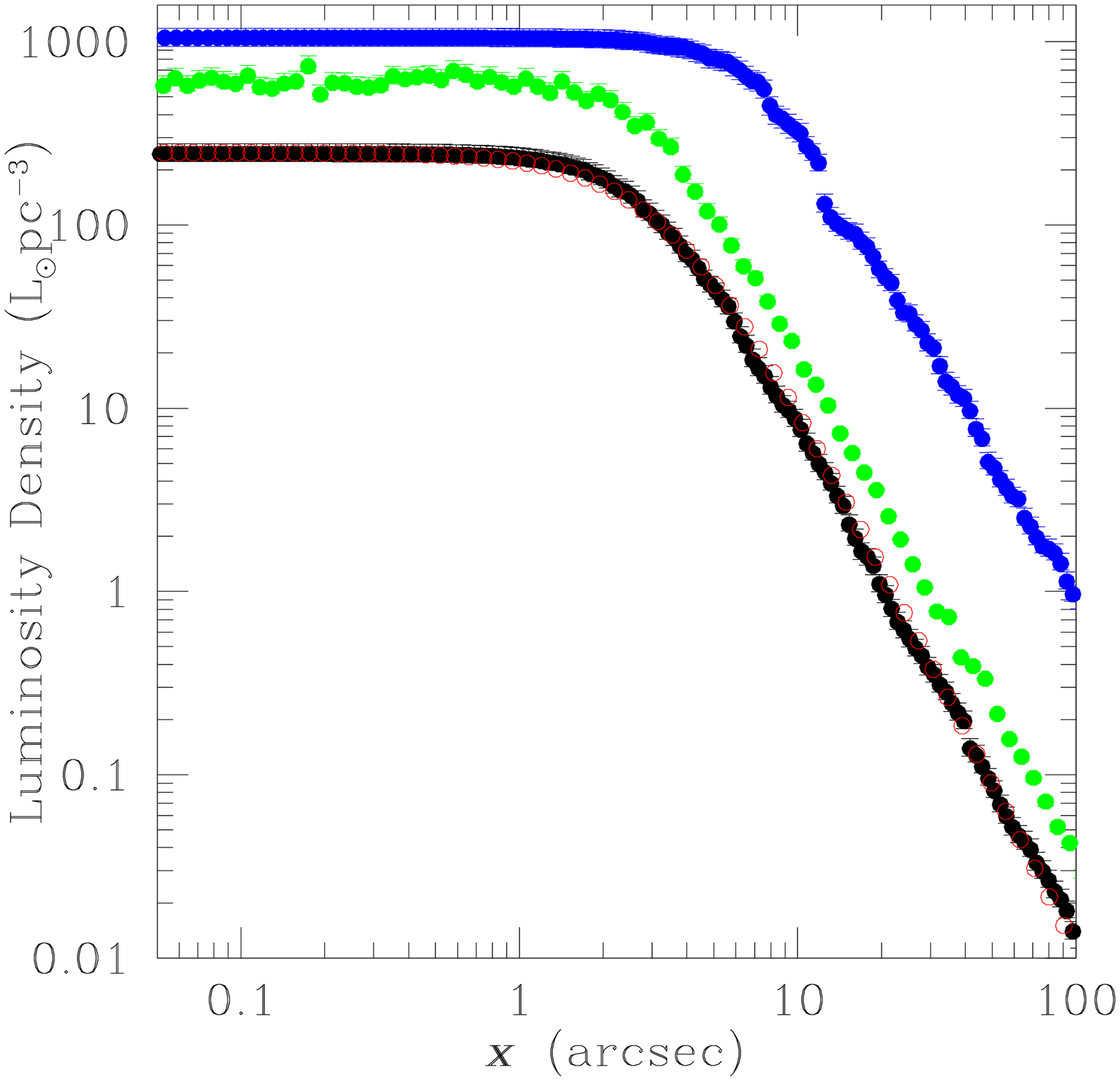}}
  \caption{Luminosity density of our oblate test galaxy of projected
  eccentricity 0.99 (shown in red), recovered by DOPING, under the
  assumptions of prolateness (in blue), oblateness (in black) and
  triaxiality with ratio between LOS extent and photometric major axis =
  0.5 (in green). All the deprojections were carried out for an
  edge-on viewing.
  \label{fig:prolate}}
  \end{figure}

  \noindent
  In the last section we explored how our unfamiliarity with the
  inclination of a galaxy can lead to a non-zero range of possible
  density profiles that a single observed brightness profile can
  correspond to. This range had been investigated under an assumption for the
  geometry of the galaxy, namely oblateness. In this section, we attempt
  to gauge the effects of treating an intrinsically oblate system, (our
  test system of Equation~\ref{eqn:density}, conferred a constant $e_p$
  of 0.99) as triaxial (with the photometric major axis along
  ${\bf{\hat{x}}}$ and LOS extent set to half the photometric major axis
  ), prolate and spherical, viewed at $i$=90\deg~(see
  Figure~\ref{fig:prolate}).

  Assuming this rather flat test system to be oblate implies that $q_1$
  is a constant, =1 and $q_p = 1/\sqrt{1-e_p^2} \approx 7.1$ (where we
  have used our definition of $q_1$ and $q_p$, as given in
  Appendix~C). Then from Equation~\ref{eqn:ep} we get that for
  $i$=90\deg, $q_2 \approx 7.1$.  Similarly, when the system is assumed
  prolate, $q_1$=1, $q_p = \sqrt{1 - e_p^2} \approx 0.14$ and $q_2
  \approx 0.14$. When we assume the system to be triaxial as above, then
  for $i$=90\deg, $q_{LOS}$=0.5 and $q_p\approx$7.1, $q_2\approx$7.1 and
  $q_1$=2.

  When we input these different values of $q_2$ in Equation~5, we get values
  of $z_{max}^{jk}$ from the oblate case that are different from what we
  get for the prolate case. In fact, for edge-on viewing, as in here,
  for a given $k$, the maximum $z$-height attained by any point in the
  $k^{th}$ isophotal annulus is higher for the oblate case than the
  prolate case. As a result, the density distribution that is recovered
  from the oblate case is lower in amplitude than that from the prolate
  case. The triaxial case result falls in between that from the oblate
  and prolate cases.

  In the case of galaxy clusters, when $q_{LOS}$ information is available,
  DOPING can be called in to perform deprojection in the fully triaxial
  geometry without requiring to make any assumption about one of the intrinsic
  axial ratios (Chakrabarty, de Filippis and Russell, 2008). 

  \subsection{Effect of PSF}
  \label{sec:psf}
\begin{figure*}
  \hskip-.01cm{
  \includegraphics[scale=.85]{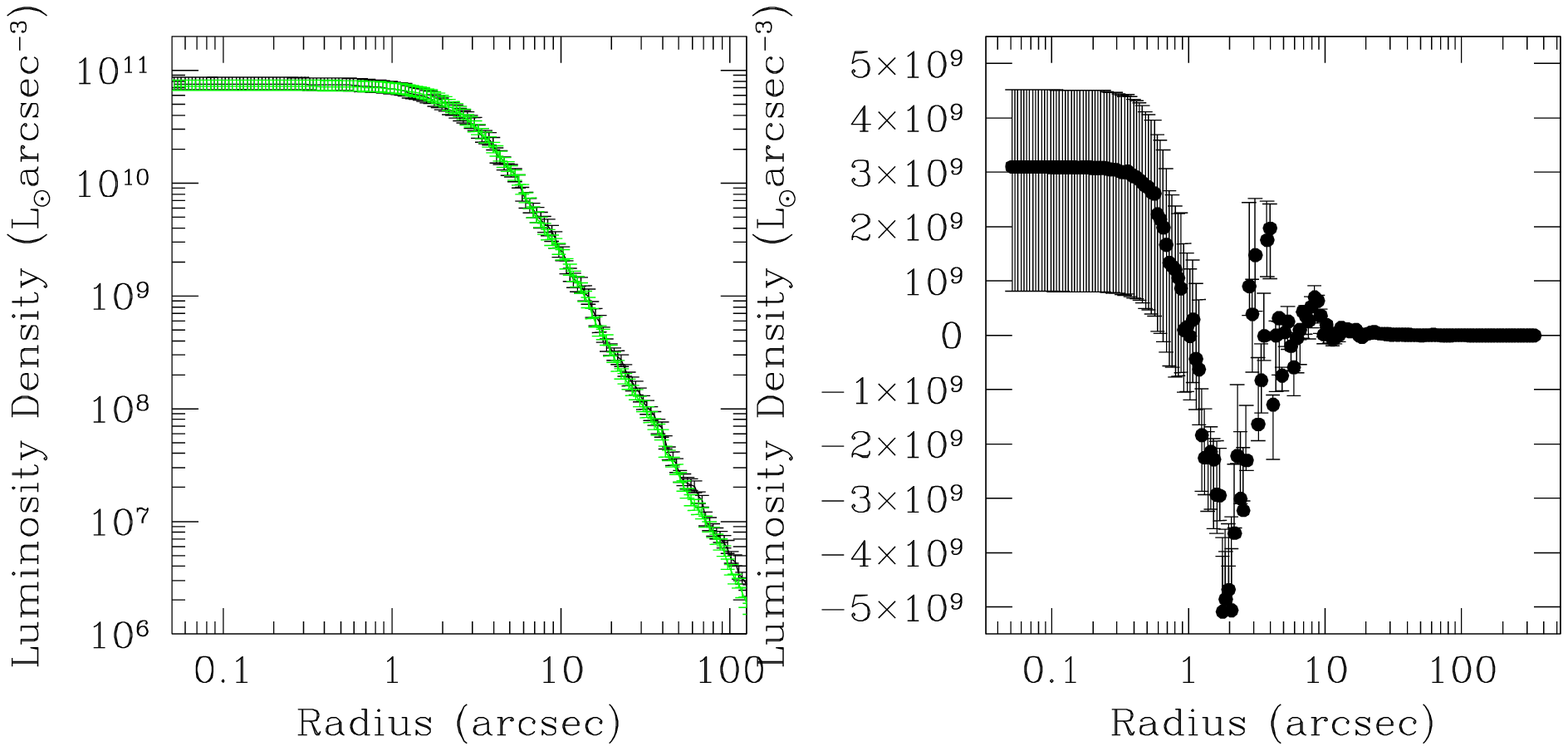}}
  \caption{Left: luminosity density of an oblate test galaxy with
  uniform eccentricity of 0.99, recovered by comparing the input
  brightness profile with the PSF convolved projection of the density
  in any iterative step. The PSF in question is the ACS PSF in the
  F850W filter. When the convolution with the PSF is ignored, the
  recovered density is shown in green. Right: difference between the
  density profiles obtained with and without convolving with the
  PSF. It is noted that inside the central 10$^{''}$, this difference
  is 2 orders of magnitude less than density while outside 10$^{''}$,
  the difference tends to zero.}
  \label{fig:psf}
  \end{figure*}

  \noindent
  We hope to use DOPING to extract luminosity profiles of real galaxies,
  in particular, the ACS VCS galaxies. It therefore becomes important to
  gauge the effect of the ACS PSF on the recovered density. This is done
  by convolving the projection of the density in any iterative step with
  the ACS PSF and comparing this convolved profile to the observed
  surface brightness. The result is shown in Figure~\ref{fig:psf}. As indicated in the
  figure, the effect of the ACS PSF does not extend beyond the central
  few arcseconds (in fact, 10$^{''}$).

\section{Applications to real systems}
\noindent
In this section, we demonstrate the application of DOPING to real
galaxies.  The efficacy of DOPING in dealing with galactic systems
that vary over wide ranges of magnitudes and morphology - including a
nucleated disky galaxy - is advanced with applications made to the
observed galaxies Ic~3019 and Ic~3881. In addition, the recovery of
the density for the cluster A1413, without resorting to assumptions
about geometry and inclination, is also included.

  \subsection{Ic~3019 - effect of smoothing}
  \label{sec:alpha}
  \noindent
  Here we apply DOPING to deproject the measured surface brightness map of the galaxy
  Ic~3019 (vcc9) which is observed within the ACS VCS \citep{laura06}.
  In particular, the effect of the smoothing parameter $\alpha$ is
  demonstrated in the context of this example galaxy. Thus, this
  section also brings out an application of DOPING to the analysis of
  real data. This galaxy is low on brightness and the reason for
  choosing it is to adduce evidence for the wide range of systems that
  DOPING can tackle.  

  The eccentricity of this galaxy has been measured to vary with radius,
  though not radically, under the ACSVCS observational program. In fact,
  eccentricity has been reported to be uniform at about 0.85 till about
  2.5\sec, from which it drops abruptly to about 0.3 at about 6\sec, to
  undulate its way up to about 0.7 at about 200\sec.

\begin{figure*}
  \begin{center}
  \includegraphics[scale=.8]{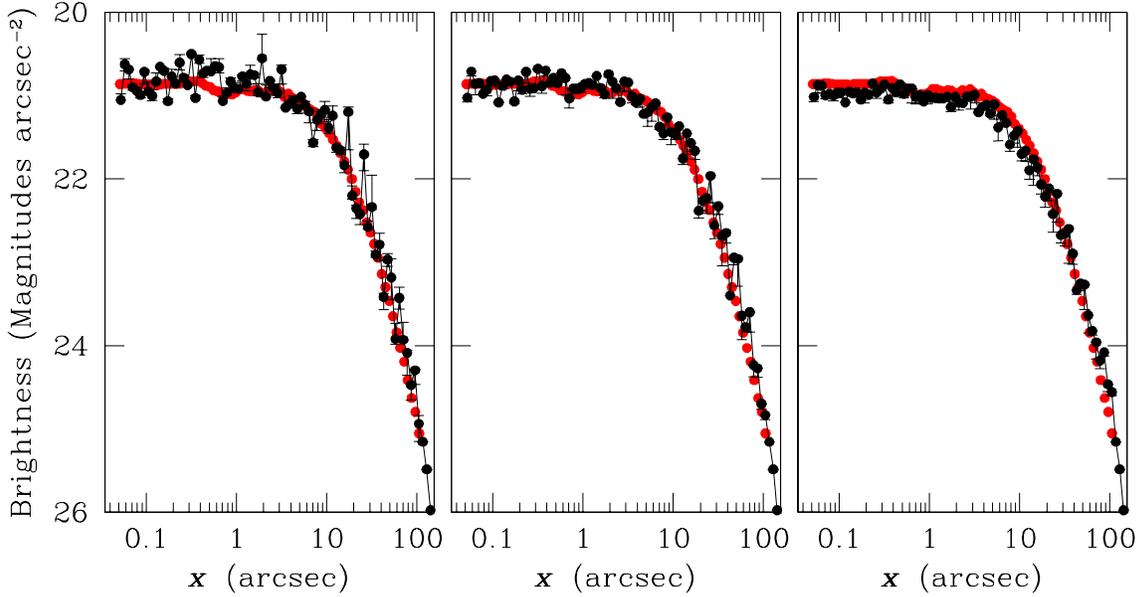}
  \end{center}
  \caption{Figure to exhibit the effect of increasing the smoothing
  parameter $\alpha$, on the recovered density and hence its projection,
  which is plotted here in black, as a function of the major axis
  coordinate, over the brightness data for I3019 (vcc9), as was
  observed within the ACS VCS program (in red). As we proceed from left
  to right, $\alpha$ goes as 0, 0.1 and 1.
  \label{fig:alpha}}
  \end{figure*}

  The density distribution recovered for Ic~3019 is projected along the
  LOS and is plotted as a function of $x$ in Figure~\ref{fig:alpha} in
  black, on top of the observed brightness data for vcc9. The three
  panels correspond to runs performed with three increasing values of
  the regularisation parameter $\alpha$, namely $\alpha$=0 (i.e. no
  smoothing), $\alpha$=0.1 and $\alpha$=1. Increasing $\alpha$ beyond
  this value did not make a significant change in the estimated density.
  The procedure to choose $\alpha$ is discussed in Appendix~E. 

\begin{figure}
  \hskip-1cm{
  \includegraphics[scale=.5]{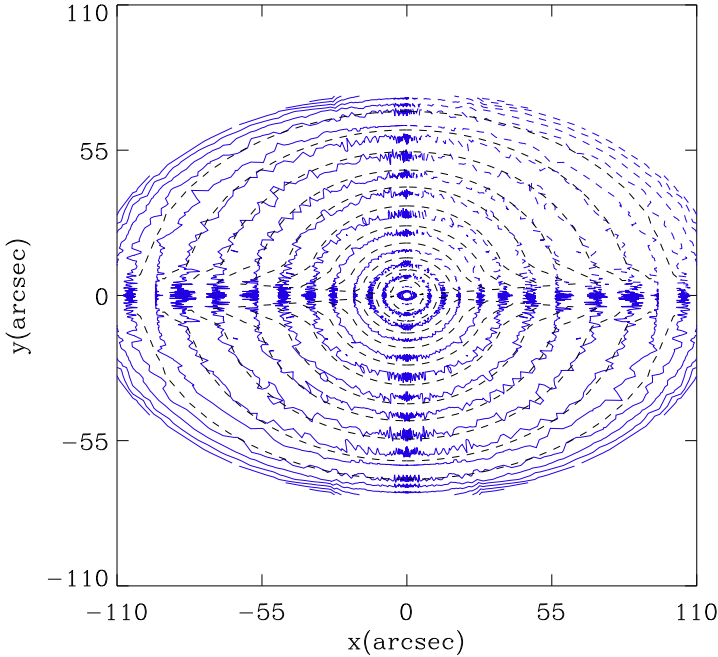}}
  \caption{As in Figure~\ref{fig:contour_rc100}, except that in this
  case, the plane of the sky brightness distribution of the galaxy IC
  3019 is shown.
  \label{fig:contour_vcc9}}
  \end{figure}

  The projection of the recovered density distribution on the plane of
  the sky, is compared to the surface brightness map of IC 3019 in
  Figure~\ref{fig:contour_vcc9}.

  \subsection{Galaxy cluster}
  \label{sec:A1413}
  \noindent
  In this section we discuss the results of applying DOPING to extract
  the X-ray luminosity density of the cluster A1413. The important
  feature about recovering the 3-D density of clusters with DOPING is
  that the true axial ratios and inclination can be constrained along
  the lines advanced by \cite{betty_08}, as long as $q_{LOS}$ of the
  system can be estimated from the available SZe measurements. The
  cluster A1413 was reported in \cite{betty_08} to be a triaxial
  system with the intrinsic axial ratios of 0.96 and 1.64 and
  inclination lying between 66$^\circ$ and 71$^\circ$. 

\begin{figure}
  \centering{
  \includegraphics[scale=.45]{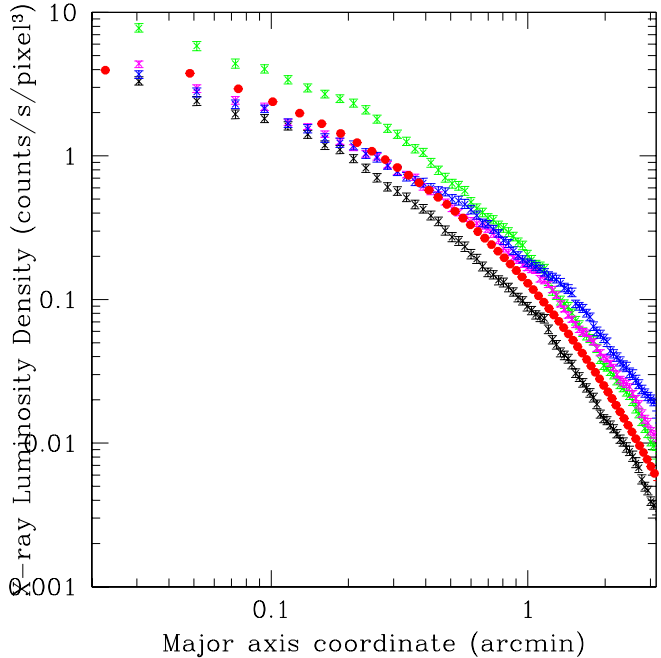}}
  \caption{X-ray luminosity density profile of cluster A1413,
  recovered under the true system geometry ($q_1$=0.96, $q_2$=1.64)
  and inclination = 68.5$^\circ$, i.e. the medial value of the
  identified range (as found by Chakrabarty, de Filippis $\&$ Russell,
  2008), is plotted in red. The other profiles mark the densities
  recovered under four distinct assumptions about the system geometry and incinations - the four benchmark deprojection scenarios (see text
  in \S~ref{sec:A1413}) used in Chakrabarty, de Filippis $\&$ Russell
  (2008).
  \label{fig:A1413}}
  \end{figure}

  This configuration was identified upon deprojecting the X-ray surface brightness
  at four benchmark deprojection scenarios, namely oblateness and
  $i=i_{min}$, oblateness and $i=90^\circ$, prolateness and
  $i=i_{min}$, prolateness and $i=90^\circ$. Here $i_{min}\approx
  47^\circ$ is the minimum inclination possible under the assumption
  of oblateness, given a projected axial ratio (= 1.473 for A1413).
  Inter-comparison of the 3-D density profiles recovered under these
  four scenarios leads us to the aforementioned prediction. Since the
  relative extent along three mutually orthogonal axes are known in
  this case, 3-D modelling is possible, i.e. the true geometry of the
  system can be estimated. Thus, we do not need to assume any axial
  ratio or inclination value. The density profile recovered under
  deprojection in the identified system geometry is presented in
  Figure~\ref{fig:A1413}.

  \subsection{2-component Galaxies}
  \label{sec:clumpy}
\begin{figure*}
  \centering{
  \includegraphics[scale=.85]{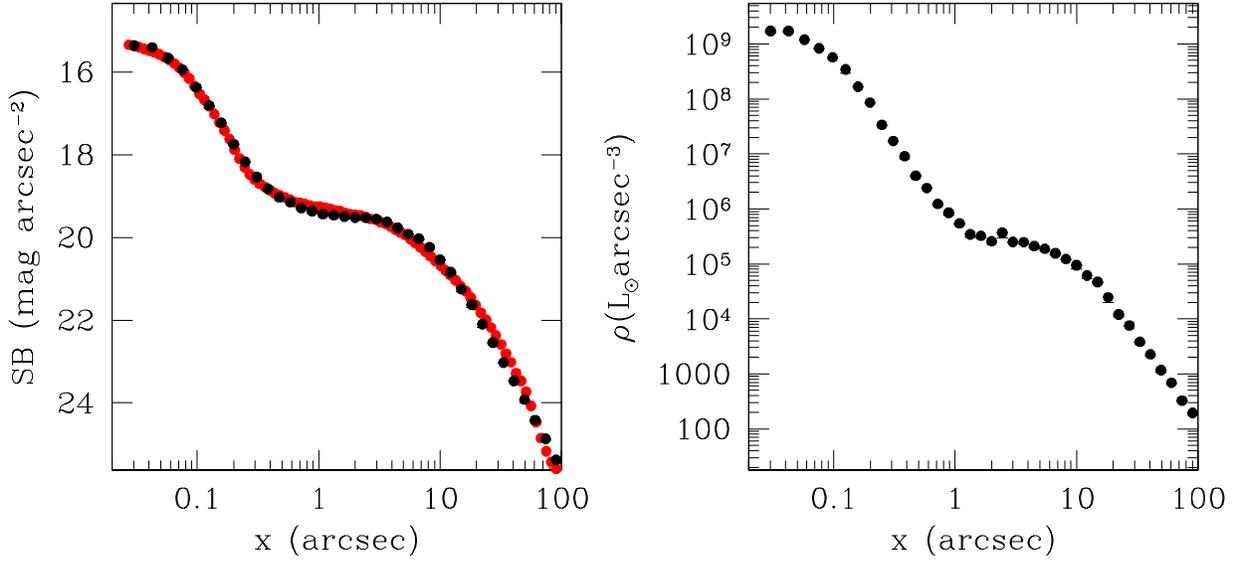}}
  \caption{Luminosity density of the two-component galaxy Ic~3881 along the ${\bf{\hat{x}}}$-axis (right) and its projection (left) in black with the observed surface brightness data superimposed in red.
    \label{fig:stuart}}
  \end{figure*}

  \noindent
  It is possible for DOPING to perform the deprojection of a bulge+disk
  galactic system in an integrated, single step fashion. This is made
  possible by ascribing two distinct seeds to the two components, namely
  the central bulge/nucleus and the more extended outer component upon
  which the central component is superimposed. The deprojection of the
  nucleated galaxies in the ACS VCS sample has been undertaken in
  Chakrabarty $\&$ McCall (2009, {\it{under preparation}}). An example
  of the deprojection of the surface brightness profile of such a 2-component, nucleated
  galaxy is shown in Figure~\ref{fig:stuart}. 

  \section{Discussions $\&$ Summary}
  \label{sec:discussions}
  \noindent
  In this paper, we have presented a fast 
  (a typical run takes about two minutes on an Intel
  Xenon 3.2GHz CPU processor) 
  new non-parametric algorithm, DOPING, that works via a penalised
  likelihood approach. It attempts to deproject the observed surface
  brightness profiles of galaxies, in general triaxial geometries, while
  taking into account intrinsic variation in shape. 

  The algorithm was successfully tested on toy galactic systems of
  varying morphologies, including an extreme system that was ascribed
  a small ellipsoidal bulge-like inner component that lay embedded in
  a highly flattened outer disk. Other experiments were best served by
  simulated galaxies with constant ellipticities. The code was also
  applied to a dwarf elliptical galaxy Ic~3019 (vcc9) and another dE,
  nucleated galaxy Ic~3381 (vcc1087) from the ACS Virgo Cluster Survey
  (C\^ot\'e et al. 2004). An application to a real galaxy cluster, A1413,
  is also included.

  \subsection{Superior Design}
  \label{sec:gielis}
  \noindent
  The well-defined, inherent convergence criterion of DOPING, buffeted
  by the sophisticated MCMC optimiser renders it superior to other well
  used inverse deprojection scheme, namely the Richardson-Lucy
  algorithm. Besides, the nonparametric inverse design of DOPING helps
  it avoid risky practises such as parametric fitting, interpolation,
  etc. Finally, none of these methods can perform deprojection under
  generalised triaxial geometries, like DOPING can.

  Choosing the LOS coordinate or $z$ as the basis for the density
  helps to keep the code modular. As a result, DOPING can handle
  deprojections in general geometries. The current version of DOPING
  relies on the determination of the intrinsic shape parameters (axial
  ratios) from measurements only of projected shape parameters. Such a
  relation is possible for unknown inclinations, only if the object
  shape bears a certain regularity. It is this that causes 3-D
  modelling with DOPING to be restricted to only objects with $m$-fold
  symmetries. 

  Thus, when the inclination is unknown but the relative extent along
  three mutually orthogonal axes are known, DOPING can handle an
  assortment of 3-D shapes that resemble $m$-winged star-fruit-like
  shapes, the 2-D projections of which are $m$-pronged star-fish-like
  2-D shapes. Such generalised geometries can be described by the 3-D
  extension of Gielis's ``superformula'' \citep{gielis}. The
  superformula is a 6-parameter generalisation of the superellipse
  (which are the Lame' curves with unequal semi-axes). In fact, the
  product of two superformulae - one corresponding to a generalised
  superellipse in the $Z$=0 plane and another in the $Y$=0 plane - can
  give rise to the $m$-winged star-fruit-like 3-D shapes\footnotemark.

  \footnotetext{Such geometries, though not relevant to astrophysics, occur in
  nature. For example, 2-D images of grown nanoparticles, taken with a
  Transmission Electron Microscope, exhibit a wide variety of shapes
  that can be accommodated by the superformula \citep{yong_06}.} 

  However, {\it a more generalised version of DOPING that allows fast
  and robust three dimensional modelling in unrestricted geometries is
  also possible - only in situations in which the system can be viewed
  at multiple inclinations}. Such configurations are not of
  astronomical context but bear strong application potential; this
  will be reported in a future contribution.

  \subsection{DOPING Deals with Substructure}
  \noindent
  Continuing on the issue of code generality, we recall that in
  \S4.2 it was shown that DOPING can deal with galaxies, the light
  distribution of which betray a bulge+disk structure. In fact,
  DOPING is capable of deprojecting systems marked with multiple
  structures that may not necessarily be concentric. As long as the
  centres of each component are known, the algorithm can be employed
  for deprojection.

  \subsection{Why Choose ${\hat{\bf X}}$=${\hat{\bf x}}$}
  \noindent
  In general, there will be two angles involved in the rotation matrix
  that connect the the two Cartesian coordinate frames. However, when
  we speak about inclinations of observed astronomical systems, we
  typically specify one angle of inclination for a given system. In
  other words, it is modelled that one of the two angles of
  inclinations is set to zero, while the angle between a principle
  axis of the system and the LOS is advanced as the inclination
  $i$. Given that the image plane is what is fixed by observations -
  and therefore the LOS - one of the system principle axes (one that
  we refer to as the $Z$-axis, say) can be at angle $i$ with the LOS,
  i.e. ${\hat{\bf Z}}$ can lie anywhere on the surface of a cone that
  has an axis along the LOS and a semi-angle $i$.  There is no
  observational constraint that can restrict our choice of the
  location of ${\hat{\bf Z}}$ on the surface of this cone. For a given
  choice of this location, the $X-Y$ plane is fixed accordingly. The
  choice that we make in this work, corresponds to ${\hat{\bf
  X}}$=${\hat{\bf x}}$.

  It is true that the recovery of the 3-D density distribution could
  be affected by a different choice for the location of ${\hat{\bf
  Z}}$ on the surface of the cone. This is so because the system at
  hand is triaxial in general, rather than axisymmetric. At the same
  time, we need to appreciate that there is no observational
  information that would inspire a particular choice. Hence we adopt
  the choice that eases calculations, keeping in mind the fact that as
  a consequence of this, the recovered 3-D density structure is one
  possible answer for a given surface brightness data. Of course, we could undertake
  deprojection for other non-zero values of $\cos^{-1}{\hat{\bf
  X}}\cdot{\hat{\bf x}}$. In fact, a band of uncertainties on the
  recovered 3-D density can be derived, corresponding to varying
  choices of this angle, though no ``most-likely'' region for the
  density can be identified within this band. However, given the state of
  equally poor constraint on the density, the solution corresponding
  to one given value of the azimuthal inclination is advanced here.

  The other assumptions that we make about geometry and inclination -
  provided by the user as inputs into the code - can be varied and the
  corresponding range of recovered 3-D density distributions can be
  recorded, with ${\hat{\bf X}}$ always assumed equal to ${\hat{\bf
  x}}$. This is particularly easy, given the short run-times of a
  typical run of DOPING. It is important to stress here that our
  assumptions are not invoked to cover for flaws in algorithm design
  but are essential in order to render the deprojection scenario
  unique, i.e.  to ascertain the deprojection geometry and
  inclination. Our assumptions merely compensate for the lack of
  (observational) information about such deprojection scenarios.

  Given the assumptions that we need to make, the question that may be
  asked is if the proposed ambition of DOPING to deproject in triaxial
  geometries is inane, in that it is driven by unconstrained
  assumptions. Such a worry has been addressed in the beginning of \S2
  and discussed later in \S~5.5. In the following section, we
  elucidate the follies of assuming axisymmetry, given observations on
  galactic systems, thus, reinforcing the need for invoking
  triaxiality.

  \subsection{The Folly of Axisymmetry}
  \noindent
  A general non-irregular galaxy, whether elliptical or axisymmetric,
  can be approximated as a triaxial ellipsoid (the third axis is tiny in
  a disk system, compared to the other two intrinsic axis lengths).  The
  modular structure of DOPING allows for deprojection of galaxies of
  both elliptical and disky morphologies. As discussed above, other
  deprojection methods can at most imply axisymmetry.

  However, often, the assumption of axisymmetry (Magorrian 1999, RK) is
  not just a mild deviation from the truth but is plain wrong - this is
  clear in the cases of inclined, disk-like systems. In these systems,
  while the extent along the perpendicular to the disk is much smaller
  than that along the other two axes in the disk, a non-zero
  projected ellipticity ($\epsilon$), measured on the plane of
  projection implies that in general, both intrinsic axial ratios -
  and particularly the ratio of the principal axes in the disk -
  deviate from unity. A few examples of such systems from the ACS Virgo
  Cluster Survey (Ferrarese\etal 2006, web site of ACS Virgo Cluster
  Survey Databases) include:
  \begin{itemize}
  \item NGC~4382 (or vcc~798, SA galaxy) in which
  $\epsilon$ ranges from 0.6 at the centre to 0.2 outside, 
  \item NGC~4762 (or vcc~2095, SB) in which $\epsilon$ approximately increases to 0.4, starting from about 0, 
  \item NGC~4442 (or vcc~1062, SB) in which $\epsilon$ increases outward to 0.6 from about 0, 
  \end{itemize}
  These examples bring out the fact that axisymmetry is not an
  acceptable approximation but an excuse for a scheme that fails to
  incorporate an improved alternative. Of course, in lieu of information
  along the LOS, even under the assumption of axisymmetry, some
  assumption has to be made about the intrinsic axial ratio. If the
  inclination permits a good estimate of the same, then no such
  assumption needs to be resorted to, whether with assumed axisymmetry
  or with DOPING.

\subsection{Justification of Assumptions}
\noindent
The solution that we will achieve for the most likely 3-D density,
  given the surface brightness data, will depend on the assumptions that we use for
  the unconstrained inclination and the axial ratio. Of course, such
  assumptions will call for physical justification - however, the
  fundamental issue here is that for galaxies, {\it there is no
  observational evidence that will constrain such assumptions}. For
  galaxy clusters, the availability of the maximum extent along the
  LOS, (from SZe measurements), definitely helps to constrict the
  number of assumptions that we then need to make (see
  Appendix~B). Similarly, for systems which can be viewed at selected
  inclinations, 3-D modelling is rendered robust and fast. Again, for
  flattened systems, we can estimate one of the inclinations and
  therefore deprojection then entails one less assumption. However,
  the lack of sufficient physically relevant measurements means that
  assumptions invoked to characterise a general triaxial system cannot
  be justified to satisfaction. In lieu of this, all that DOPING can
  offer is a fast estimate of the range of density distributions
  obtained over the considered range of axial ratios and
  inclinations. Turning this argument around, we realise that the
  range of 3-D density distributions that are recovered in \S3.2 and
  3.3, for distinct geometry+inclination inputs (assumptions) cannot
  be narrowed down for general triaxial galactic systems.

  The question that then begs addressing is if deprojection in triaxial
  geometries is any improvement upon the existing deprojection routes
  that are currently in vogue. That the need for triaxiality over
  axisymmetry, is physically justified, was delineated in the last
  sub-section. However, given the dearth of observational information
  - particularly for galaxies rather than galaxy clusters, as
  discussed in \S~2 and above - assumptions need to be invoked. The superiority
  of DOPING lies in the fact that when information about intrinsic
  morphology and inclination are less sparse than for galaxies (as for
  clusters or deposited nano-particles), identification of the true
  triaxial geometry is possible \citep{betty_08} and deprojection can
  then be performed in this geometry, without invoking assumptions
  about $i$ and $q_1$ (as demonstrated for the Abell cluster A1413 in
  \S~\ref{sec:A1413}). 

  \subsection{Position Angle}
  \noindent
  Had the two angles of inclination been known to us, we would be in a
  position to predict the observed projected position angle as a
  function of these inclinations. However, given the projected
  position angle, constraining the inclinations requires an inverse
  approach, which is possible within DOPING in the triaxial
  geometry. We could then relax the assumption of ${\hat{\bf
  X}}$=${\hat{\bf x}}$ while continuing to assume the polar
  inclination angle. In fact, the coordinate system (and the ensuing
  equations) used here is a limiting case of the assumption of a
  non-zero, radially varying position angle. In this more generalised
  version of DOPING, the angles between $\hat{\bf x}$ and the line of
  nodes will be non-zero and varying with radius
  \citep{simonneau98}. This will be dealt with in a future
  contribution.

  When the position angle is included in the calculations the data table
  is enhanced by another column yet - $\phi_{0k}$, where $\phi_{0k}$ is
  the position angle of points in the $k^{th}$ isophotal annulus. Then,
  the body coordinate system corresponding to the $k^{th}$ isophotal
  annulus is rotated by $\phi_{k0}$ with respect to the $\phi$=0 line,
  i.e. the $x$-axis (by definition), so that the new body coordinate
  system governing the triaxial shell - the largest projection (of the
  same thickness as itself) of which is the $k^{th}$ isophotal annulus -
  is given by $X'-Y'-Z'$, where:
  \begin{eqnarray}
  X' &=& X\cos\phi_{0k} + Y\sin\phi_{0k}\\
  Y' &=& X\sin\phi_{0k} - Y\cos\phi_{0k}\\
  Z' &=& Z
  \end{eqnarray}
  Having established this, the equivalent of Equation~\ref{eqn:ell_i}
  can be written down. The reformed equation is still a quadratic and is
  solved for the two solutions of $z$ as before. 
  In the present calculations, we avoided this extra complication in
  light of the small isophotal twist observed with the ACS VCS galaxies,
  which are the prime targets of the discussed code.

  \subsection{Effect of Seed Selection}
  \noindent
  Our starting luminosity density is motivated by crude estimates of
  the sought function \citep{gelman_book}; we are guided by the
  requirement that the projection of the seed density be close to the
  given surface brightness data. The algorithm will indeed fail to converge for
  completely irrational choices of the initial parameters (steepness
  parameter $\gg 1$, scale length different from the correct choice by
  more than 4 orders of magnitude). Importantly, under such
  circumstances, the projection of the recovered density will be found
  to deviate from the brightness data. Additionally, it may be
  remarked that it is reasonable to start with a steepness parameter
  that is close to unity and a scale length that is of the order of
  the core radius that characterises the surface brightness profile of
  the system. Given that a typical run takes less than 2 minutes on an
  Intel Xenon 3.2GHz processor, it is feasible to restart the
  algorithm for a different choice of the initial guess, until
  convergence is reached.

  \subsection{Effect of Data sampling}
  \noindent
  We also note that the spatial sampling of the data can have some
  effect on the density distribution recovered by DOPING. For instance,
  in Figure~\ref{fig:change}, the density profile best reproduces
  the data at small, rather than large radii. This is a consequence of
  the fact that, by construction, our simulated data set happens to have
  substantially more data points in the inner regions of the galaxy than
  on the outside, with the consequence that the innermost region has
  larger weight in driving convergence.

  \section{Conclusion}
  \noindent
  Thus, DOPING is advanced as a simple but powerful deprojection
  algorithm, that can be treated as a black-box by the user, is
  fast and offers 3-D density distributions in general geometries,
  without resorting to making unconstrained approximations to the form
  of the density or blindly accepting validity of commonly used
  goodness of fit measures in light of the inhomogeneous errors of
  measurement \citep{bissantz_01}.

  The greatest novelty borne by DOPING is its applicability to general
  systems. Even though in the above examples, triaxial systems were
  investigated - ranging from razor thin discs to 2-component or
  bulge+disk galaxies - even when inclination is unknown DOPING can
  deal with all systems that offer an analytical relation between the
  intrinsic and measured projected shape parameters. The class of
  geometries that bear an $m$-fold symmetry allows for this. 3-D
  modelling of images of systems with even more general geometries can
  also be performed with DOPING, as long as the system can be imaged
  at various known inclinations. That deprojection into the third
  dimension is possible in such general geometries, in a
  non-parametric way, is due to the fact that the representation of
  the sought density is modular, i.e. not dependent on a
  characteristic of the system geometry.

  The recovery of the 3-D density is performed iteratively, by
  searching for the most likely 3-D density structure that projects to
  the observed 2-D image. This search is robustly undertaken by an
  MCMC optimiser. Since the choice of the 3-D density is constrained
  via its projection, distinct 3-D density distributions will project
  to the same 2-D image. To lift this degeneracy, in DOPING, the
  system geometry and orientation are specified completely. That
  axisymmetry is an invalid assumption - at least in real disc-like
  galaxies - was shown above. Consequently, the description of galaxy
  geometry as triaxial is a suitable alternative. However, triaxiality
  entails two axial ratios and inclinations, not all of which can be
  specified, given the constricted level of achievable observed
  information. When observed information is available, DOPING can
  perform deprojection under triaxiality without invoking assumptions,
  while in lieu of the same, assumptions are invoked. The former case
  is demonstrated above via the example of deprojection of a galaxy
  cluster. A benefit of the speed of the algorithm is that a suite of
  3-D density models, corresponding to a range of assumed values, is
  achieved quickly.

  Thus, the strengths of DOPING include generalised applicability,
  ability to incorporate substructure and non-parametric density
  recovery, along with logistical advantages such as high speed of
  runs and user-friendliness. Such characteristics render DOPING a
  very useful tool in three dimensional modelling. In particular, the
  all important estimation of galactic masses will be aided by a tool
  such as DOPING.

  \appendix{
\section{Representing isophotes}
\noindent
The shape parameters of the isophotes form the input information, so
we can formulate smooth analytical approximations to the isophotes. Of
course, it is the fitting of such smooth approximations to the surface
brightness data that provides estimates of the isophotal parameters;
in other words such approximations are readily available. It is to
sort grid points on the image plane, into respective isophotal annuli,
that we invoke these approximations. However, real isophotes can be
irregular and not altogether smooth. To take this into account, we
examine the isophotes first and estimate the typical length scale over
which the irregularity in the isophotes occurs. Thus, for example, the
isophotal contours of a distant galaxy could be imaged by a given
instrument as more jagged than those of a nearby system. We discard
grid points on the image plane that lie within this estimated spatial
range corresponding to deviations from smoothness.

\section{Specifying the system geometry for triaxial galaxies}
\noindent
The specification of 
triaxiality of an example galaxy entails knowledge of:
\begin{itemize}
\item 2 constant intrinsic axial ratios $q_1$ and $q_2$ for systems
with radially independent shape. Alternatively, for systems with
radially varying intrinsic eccentricities - 2 intrinsic axial ratios
that vary as known functions of distance away from centre of system.
\item 2 position angles or inclinations.
\end{itemize}
Of these, we
\begin{itemize}
\item set one inclination angle to 0 by setting one photometric axis
(along the ${\hat{\bf x}}$-axis) to be coincident with an
intrinsic principal axis (see \S~2.1).
\item assume a value for the other inclination $i$.
\item derive the two intrinsic axial ratios from the two projected
axial ratios: 
\begin{enumerate}
\item ratio of the photometric semi-axes ($q_p$).
\item ratio of the semi-axes along the LOS to that along the
photometric major axis ($q_{LOS}$).
\end{enumerate}
\end{itemize}
However, generally we cannot measure the extent of a galaxy along the
LOS. This is not so for other systems though, eg. galaxy clusters, the
LOS extent of which is constrained by SZe measurements; for such
systems, we can solve for $q_1$ and $q_2$ from:
\begin{itemize} 
\item $q_p=f_1(q_1, q_2, i)$ and
\item $q_{LOS}=f_2(q_1, q_2, i)$,
\end{itemize} 
where $f_1, f_2$ are analytical functions whose forms can be derived
from geometrical considerations. In lieu of the measure of the system
along the LOS, we {\it {compensate for our ignorance of $q_{LOS}$ by
assuming one of the intrinsic axial ratios}} ($q_1$). Then using
this assumed $q_1$, we can obtain $q_2$. With $q_{LOS}$ available,
deprojections of X-ray brightness data of clusters have been performed
by DOPING in triaxial geometries, without requiring any assumption for
$q_1$ (Chakrabarty $\&$ de Filippis, {\it{under preparation}}).

\section{Axial ratios}
\label{sec:axial}
\noindent
We clarify that the axial ratios $q_{2k}, q_{1k}, q_{pk}$ are defined
such that the extent along $\hat{\bf x}$ is in the numerator. Thus,
\begin{itemize}
\item $q_{pk}$ is the ratio of the semi-axis along $\hat{\bf x}$ of the $k^{th}$ isophotal annulus, to that along $\hat{\bf y}$, on the image plane. 
\item $q_{2k}$ is ratio of the semi-axis along $\hat{\bf x}$ to that along the $\hat{\bf Y}$.  
\item $q_{1k}$ is ratio of the semi-axis along $\hat{\bf x}$, to that along the $\hat{\bf Z}$. 
\end{itemize}
where $\hat{\bf x}=\hat{\bf X}$ (see \S~2.2). 

In the examples shown later in the paper, we assume the system to be
oblate, unless otherwise stated, i.e. then $q_{1k}$ is a constant,
(=1) and $q_{2k} >$ 1. If we are considering a prolate system, then
$q_{1k}$ is again a constant, (=1) but then $q_{2k} <$ 1. 

  \section{Optimisation}
  \noindent
  We seek solutions for $\rho$ that correspond to the $\pm$1-$\sigma$
  neighbourhood of the global maxima of ${\cal{L}}$ and employ an MCMC
  optimiser for this. The particular implementation of this in our
  work is the Metropolis Hastings algorithm. Once Metropolis-Hastings
  attains the equilibrium stage, it moves around in the maximal region
  of ${\cal{L}}$. During this stage the average of an ensemble of
  models should represent the distribution that is to be sampled; this
  is reflected in stationarity in the trace of the likelihood. Before
  recording the solutions, we typically allow for 5 times the period
  of burn-in to lapse, mindful of the fact that burn-in can continue
  much longer than suggested by the trace. Details of the optimisation 
  are presented in \cite{gelman, roberts_97a, roberts_97b, roberts_01}.

  We use circular iterations, with the 1$^{st}$ to the $N_{iter}^{th}$
  iterative step repeating in cycles. We define convergence as when
  inside the equilibrium stage, the likelihood attained in the $i^{th}$
  step during the $M^{th}$ cycle, falls below the likelihood attained in
  the $i+1^{th}$ step during the $M-1^{th}$ cycle. It is expected
  that then the algorithm has indeed passed through the global maxima
  in the likelihood. The extent of wandering of Metropolis in this
  maximal region is a direct measure of the errors of the analysis.

  \subsection{Uncertainties}
  \noindent
  In fact, when we say that the errors of our analysis are quantified
  by the $\pm$1-$\sigma$ spread across the ensemble of identified
  density values, we actually imply a as 68$\%$ interval. To expound
  on the procedure of quantifying the errors: at a point $(x,y,z)$,
  the values of the luminosity density corresponding to the maximal
  region of the likelihood function are recorded. This vector of
  density values is sorted and values at the 16$^{th}$, 50$^{th}$ and
  84$^{th}$ centiles are noted. The interval estimate of the
  luminosity density at this given point $(x,y,z)$ is then represented
  by the error band bound by values corresponding to the 16$^{th}$ and
  84$^{th}$ centiles; the density value at the medial position within
  this interval is also shown within this error band.

  \subsection{Updating Density}
  \label{sec:update}
  \noindent
  At the beginning of every step, each $\bf{z-histogram}$ is varied,
  {\it independently of each other}, while maintaining the realistic
  conditions of positivity. In general, the old value of the relevant
  variable ($X$) in step $i$ is related to a new value $Y$ in the next
  step: $Y = X + Z$, where $Z$ is a random variable. A variety of
  proposals to move from $X$ to $Y$ are described in the literature
  \citep{mengersen, gelman}.  Of these, we choose that the algorithm
  proposes to move from $X$ to $Y$ using the random walk jumping
  distribution, i.e. $Y=X + sU$, where $U$ is a Gaussian random
  deviate and $s$ is the scale parameter that determines the size of
  the jump. If the amount of change is very small, the chain will
  require a very large number of steps to become well-mixed and hence
  efficiency will be compromised. If the step-size is too large, the
  worry is that the resulting configuration will miss the global
  maximum and fall into regions of very low likelihoods, wasting a
  number of steps in the process \citep{draper_2000}. Our chosen jump
  proposal is adaptive in nature since the updating of
  $\rho(\cdot)$ at a given $z$, depends on the density at that $z$.
  The details of the density updating is as follows:
  \begin{eqnarray}
  \rho_{n+1}(x_k^j, y_k^j, z_l) &=& \rho_n(x_k^j, y_k^j, z_l+\delta{z}) +
  \Delta_l R  \quad {\textrm{where}} \nonumber \\
  \Delta_l &=& \rho_n(x_k^j, y_k^j, z_l) - \rho_n(x_k^j, y_k^j, z_l+\delta{z})
  \end{eqnarray}
  $(x_k^j, y_k^j, z_l)$ in step $n$. Also, $R\sim N(0, scl_1^2)$ and
  $scl_1$ is a scale that determines the extent of the jump in the
  shape of a ${\bf z-histogram}$ between two consecutive steps. Also,
  for this $k, j$, $z_l$ is the $l^{th}$ $z$-bin and $z_{l+1} = z_l +
  \delta{z}$. This updating is done $\forall\: l: z_l\in[0,
    z_{max}^{jk}]$, $\forall\: j, k$, to accomplish change of shape of
  the ${\bf {z-histogram}}$ corresponding to the $j, k$.

  We still need to make a change to the overall amplitude of the new
  density distribution. This is done by scaling $\rho_{n+1}(x_k^j,
  y_k^j, z_l)$ by another random deviate $\sim N(0, scl_2^2)$,
  $\forall\: l$; here $scl_2$ is another scale that determines the
  amount of change of amplitude that is brought about in the density
  structure, for a given $j, k$.

  The values of $scl_1$ and $scl_2$ are arrived at experimentally, keeping
  the effect of large and small scales in mind.

  We choose to work with equal binning along all three coordinate axes;
  this binning is logarithmic in nature. A good choice for the smallest
  bin size is of the order of the spatial resolution of the data. A typical run takes about 2 minutes on an
  Intel Xenon 3.2GHz CPU processor.

  \subsection{Temperatures}
  \noindent
  The probability of accepting the proposed move from likelihood
  ${\cal{L}}(A)$ to ${\cal{L}}(A^{'})$ is discussed here. Here $A$
  generically represents the domain of the likelihood function which is
  the set of the ${\bf{z-histograms}}$ and $A^{'}$ is the new set of
  ${\bf{z-histograms}}$ to which a move has been proposed.

  Anxiety over multi-modality of the likelihood function has prompted us
  to work both with highly dispersed initial values (or seeds) to
  initiate multiple chains \citep{gelman_rubin} and also to use simulated
  annealing in a single chain. When the latter is implemented, the transmission
  probability is
  \begin{equation}
  P(A\rightarrow{A^{'}}) = min\displaystyle{
					    \left[
						  \exp{\frac{\Delta{\cal{L}}}
						      {T}}, 1
					    \right]
					   },
  \end{equation}
  where $\Delta{\cal{L}}=\ln{ {\cal{L}}(A^{'})} - \ln{ {\cal{L}}(A)}$ and $T$
  is the temperature parameter.

  We have implemented both a uniform temperature profile, as well as
  played with a cooling schedule that starts with an initial temperature
  $T_0$ which is allowed to cool down to a final value of $T_f$, over a
  step number of $N_{cool}$. In practice, we choose $T_f/T_0$ to be 0.1
  while $N_{cool}$ is typically set to double the number of steps that
  correspond to $burn-in$, as judged from traces of system
  characteristics, such as the value of the density that is recovered at
  a fiduciary location. We find that the answer depends only very weakly
  on the details of the cooling schedule.

\section{Choice of the Regularisation Parameter}
\noindent
  \citet{thomson_91} refer to the smoothing parameter as the compromise
  between ``fidelity to the data and smoothness''. While different
  methods are suggested by \citet{thomson_91} and
  \citet{titterington}, to constrain the scalar $\alpha$, we choose
  to accept a value that is achieved via an empirical implementation of
  the minimisation of the total mean-squared error. We define the total
  mean-squared error (TMSE), for a given value of $\alpha$ as:
  \begin{equation}
  \delta_\alpha = \displaystyle{\frac{{\displaystyle\sum_{i=1}^{N}(\rho_{median}^i - \rho_{-1\sigma}^i)^2}+{\displaystyle\sum_{i=1}^{N}(\rho_{median}^i - \rho_{+1\sigma}^i)^2}}{2N}}
  \end{equation}
  where there are $N$ grid points over which a density profile is
  recovered and $\rho_{median}^i$ is the density recovered at the medial
  level in the $i^{th}$ grid point, while $\rho_{-1\sigma}^i$ and
  $\rho_{+1\sigma}^i$ are the values recovered at the -1$\sigma$ and
  +1$\sigma$ error levels, respectively. We scan over a range of
  $\alpha$ values and stop increasing $\alpha$ when the smallest
  $\delta_\alpha$ is achieved. Typically, beyond a given value,
  increasing $\alpha$ maintains the corresponding value for TMSE. It is
  the smallest $\alpha$ at which this trend is noticed that is chosen as
  the $\alpha$ we use in the runs.

  \section{Choice of Seed for Density}
  \label{sec:guess}
  \noindent
  The robustness of a recursive formalism is
  reflected in the extent to which the initial guess for the solution
  is irrelevant to the final result. With this in
  mind, we undertook extensive experimentation with seed density
  distributions used by DOPING as starting guesses. For this
  investigation, we use the same simulated data set of an oblate galaxy,
  viewed edge-on, as described above: analytical density
  of Equations~\ref{eqn:density} and a constant eccentricity of 0.99.

  Typically, we use a seed density $\rho_{seed} = \rho_{seed}(\xi)$,
  where $\xi$ denotes the ellipsoidal
  radius. In fact, we choose a seed density that has a form akin to the
  Lorentzian distribution:
  \begin{equation}
  \rho_{seed}(\xi) = \displaystyle{\frac{a}{1 + \displaystyle
			    {\left(\frac{\xi}{b}\right)^{2n}}}},
  \label{eqn:lorentz}
  \end{equation}
  The three parameters in this distribution are
  \begin{itemize}
  \item the scale length $b$ which determines the width of the profile, 
  \item the central density $a$ and 
  \item an exponent $n$ that defines the steepness of the fall of the
  tail of the profile.
  \end{itemize}

  A reasonable starting value of $n$ is about 1. We find that of these
  three parameters the algorithm is least sensitive to the choice of the
  amplitude $a$. When the scale-length and the steepness parameter are
  widely varied, the algorithm also follows the input data, though it is
  more difficult for the code to do so for radical changes in shape than
  in amplitude. To test the robustness of DOPING to the initial guess,
  for each input parameter, we performed two runs with values chosen to
  be at opposite ends of the true value. The density profile that DOPING
  retrieves when $a$ is varied, is compared in Figure~\ref{fig:change}
  to the known density distribution. The projections of the density
  profiles are also shown, superimposed on the brightness data that
  serves as the input to the code.
  \begin{figure*}
  \centerline{
  \includegraphics[width=0.8\hsize]{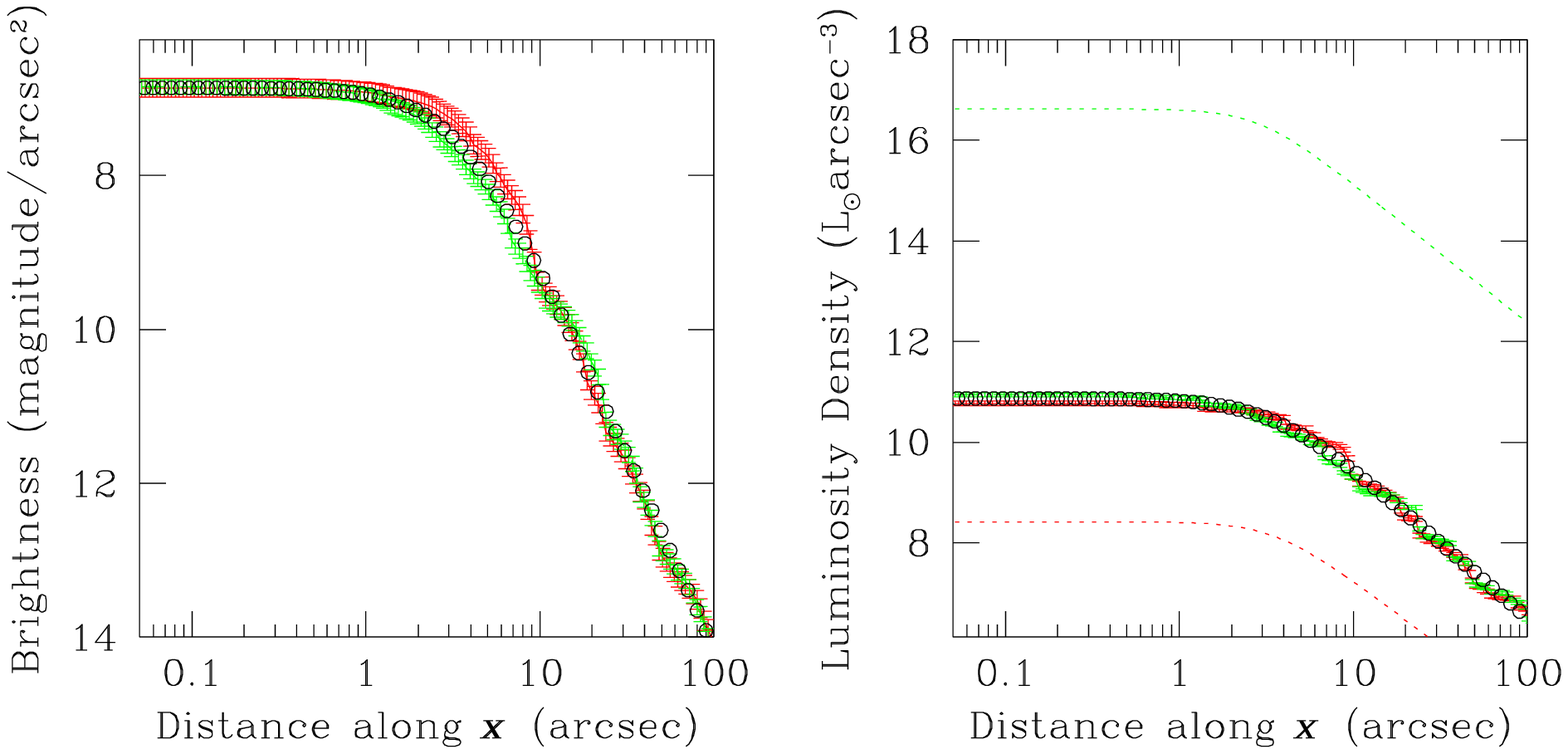}}
  \centerline{
  \includegraphics[width=0.8\hsize]{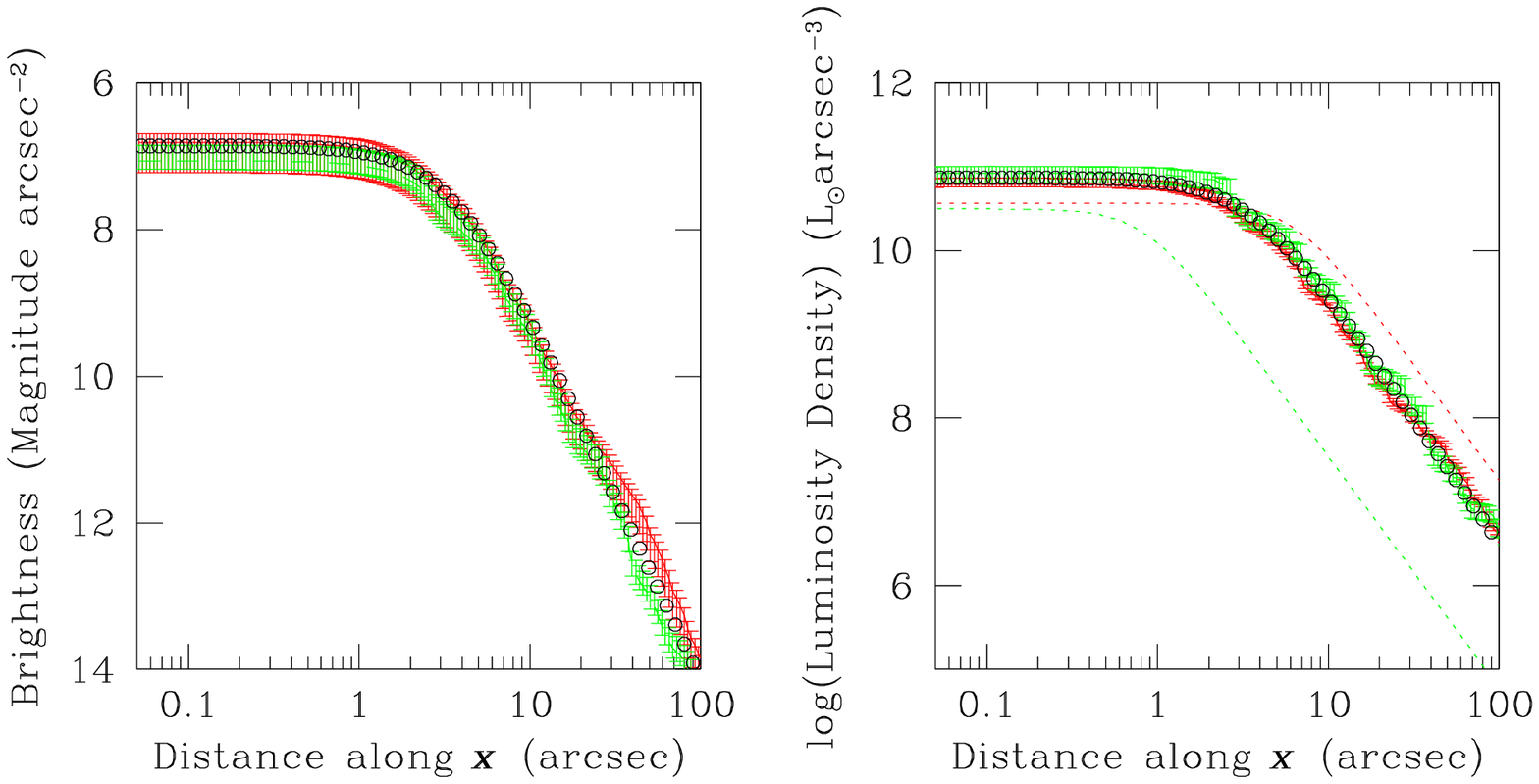}}
  \centerline{
  \includegraphics[width=0.8\hsize]{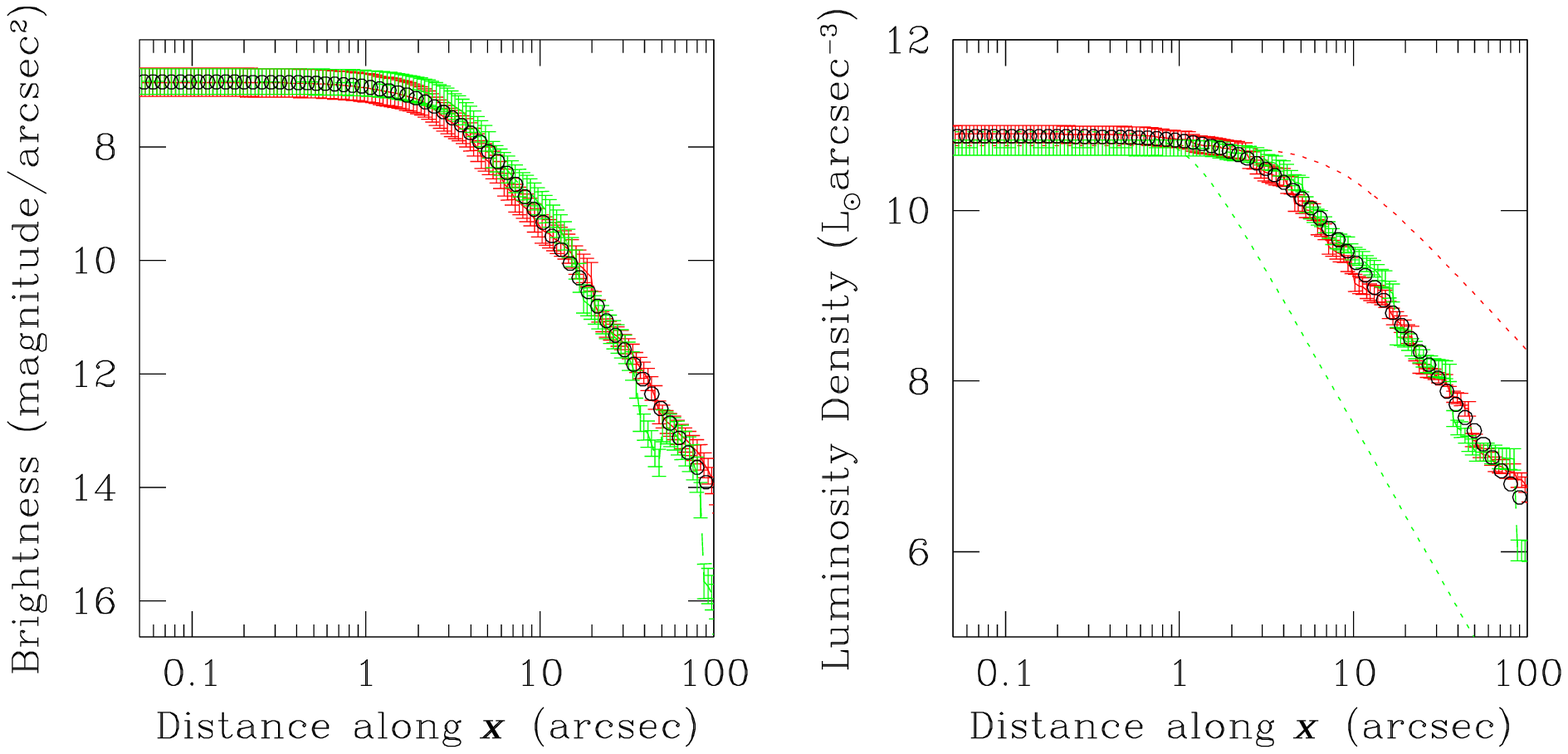}}
  \caption{Top row: figure to verify the robustness of DOPING to changes
  in the central density of the initial guess for the density
  distribution. Density profiles from two runs (in red and green,
  superimposed with error-bars), performed with values of the central
  density parameter that are apart by eight orders of magnitude, are
  shown in the right panel. Superimposed on these is the true density of
  the system, shown in black open circles. The initial seeds for the
  density distributions in the two runs are shown in dotted lines, in
  the corresponding colours. The left panel shows projections of the
  recovered density profiles in corresponding colours, with the input
  surface brightness data over-plotted as open circles. Middle row: the
  scale length parameter of the initial guess is changed. The two runs
  correspond to $b$=2\sec.5 (in red) and $b$=40$^{''}$ (in green).
  Lower row: the steepness parameter of the initial guess is
  changed. The two runs correspond to $n=1.1$ (in red) and $n=1.8$ (in
  green).}
  \label{fig:change}
  \end{figure*}

  The top panels of Figure~\ref{fig:change} shows the projected surface
  brightness and three dimensional luminosity density profiles obtained
  from runs done with initial guess characterised by values of central
  density ($a$) that are 8 orders of magnitude different. While one of
  the runs (in solid red lines) was started with a central density of
  about 3.2$\times 10^8$L$_\odot$/arcsec$^{3}$, the other (shown in
  dotted green lines) corresponds to $a \approx 3.2\times 10^{16}$
  L$_\odot$/arcsec$^{3}$. The run shown in Figure~\ref{fig:testcase} was
  carried out with $a \approx 10^{11}$L$_\odot$/arcsec$^{3}$.  

  The panels in the middle row of Figure~\ref{fig:change} display the
  recovered density profiles and their projections when the initial
  guess is characterised by scale lengths (the parameter $b$ discussed
  above) of 2\sec.5 and 40$^{''}$. These values were chosen at two
  opposing ends of the value of $b$=10$^{''}$, which was used in the
  run, the results from which are shown in
  Figure~\ref{fig:testcase}. It
  appears that in spite of the shape of the initial guess being
  significantly different from the true profile, the algorithm converges
  to the true density profile.

  In the lower panel of Figure~\ref{fig:change}, results obtained from
  runs done with a steepness parameter $n=1.8$ (in dotted red lines) and
  $n=1.1$ (in solid green lines) are shown; the plots in
  Figure~\ref{fig:testcase} were retrieved from a run done with
  $n=1.4$. The recovered density profiles from these runs are consistent
  with the true density distribution, within error bars.  The
  implications of these results are discussed in full details in
  \S~\ref{sec:discussions}.

}

\begin{acknowledgements}
\noindent
This research was funded by a Royal Society Dorothy Hodgkin
Fellowship. The author is delighted to acknowledge the contribution
of Laura Ferrarese without whose comments and suggestions, the paper
would not have been possible.

\end{acknowledgements}


\clearpage


\begin{thebibliography}{}

\bibitem[Bendinelli(1991)]{bendinelli}
Bendinelli, O., 1991, ApJ, 366, 599.

\bibitem[Bissantz \& Munk(2001)]{bissantz_01} 
Bissantz, N., \& Munk, A. 2001, \aap, 376, 735. 

\bibitem[van den Bosch(1997)]{vandenbosch}
van den Bosch, F. C., 1997, \mnras, 287, 543.

\bibitem[Cappellari(2002)]{cappellari}
Cappellari, M., 2002, \mnras, 333, 400.

\bibitem[Chakrabarty, de Filippis $\&$ Russell(2008)]{betty_08} Chakrabarty, D., de Filippis, E., \& Russell, H., 2008, \aap, 487, 75. 

\bibitem[Chakrabarty $\&$ Ferrarese(2008)]{ijmp}
Chakrabarty, D.; Ferrarese, L., 2008, {\em IJMP(D)}, 17, 195.

\bibitem[C\^ot\'e et al.(2004)]{coteacs04}
C\^ot\'e, P., Blakeslee, J. P., Ferrarese, L., Jordan, A., Mei, S., Merritt, D., Milosavljeviϙʤ, M., Peng, E. W., Tonry, J. L., West, M. J., 2004, \apjs, 153, 223.

\bibitem[Draper(2000)]{draper_2000}
  Draper, D., 2000, {\url{http://citeseer.ist.psu.edu/339805.htm}}

\bibitem[Fabricant, Rybicki $\&$ Gorenstein(1984)]{fabricant84}
Fabricant, D.; Rybicki, G.; Gorenstein, P., 1984, \apj, 286, 186

\bibitem[Ferrarese et al.(2006)]{laura06}
Ferrarese, L., C\^ot\'e, P., Blakeslee, J. P., Mei, S., Merritt, D., West, M. J., 2006, {\em arXiv:astro-ph/0612139}.

\bibitem[Gebhardt et al.(1996)]{gebhardt96}
Gebhardt, K., Richstone, D., Ajhar, E. A., Lauer, T. R., Byun, Y., Kormendy, J., Dressler, A., Faber, S. M., Grillmair, C., \& Tremaine, S., 1996, \aj, 112, 105.

\bibitem[Gelman, Roberts $\&$ Gilks(1996)]{gelman} 
Gelman, A., Roberts, G., O. \& Gilks, W., R., 1996, 
{\em{Bayesian Statistics 5}}, ed. J. Bernardo et al., 599, Oxford University Press.

\bibitem[Gelman et. al(1995)]{gelman_book} 
Gelman, A., Carlin, J., Stern, H., \& Rubin, D., 1995, {\em Bayesian Data Analysis}, Chapman and Hall

\bibitem[Gelman $\&$ Rubin(1992)]{gelman_rubin} 
Gelman, A. \& Rubin, D. B., {\em Statistical Science}, {7}, 457.

\bibitem[Gerhard \& Binney(1996)]{binneygerhard}
Gerhard, O. E., \& Binney, J. J., 1996, \mnras, 279, 993.

\bibitem[Gielis(2003)]{gielis}
Gielis, J., 2003, {\em American Jl. of Botany}, 90, 333.

\bibitem[Hastings(1970)]{hastings} Hastings, W. K., 1970, {\em Biometrika}, 57, 97.

\bibitem[Haykin(2008)]{haykin_08} Haykin, S. S., 2008, {\em Neural Networks and Learning Machines}, (Prentice Hall).

\bibitem[Jedrzejewski(1987)]{jedrzejewski}
Jedrzejewski, R. I., 1987, \mnras, 226, 747.

\bibitem[Jedrzejewski, Davies $\&$ Illingworth(1987)]{jedrzejewski87}
Jedrzejewski, R. I., Davies, R. L., Illingworth, G. D., 1987, \aj, 94, 150.

\bibitem[Kochanek \& Rybicki(1996)]{kochanekrybicki}
Kochanek, C. S., \& Rybicki, G. B., 1996, \mnras, 280, 1257.

\bibitem[Krajnovic et al.(2004)]{krajnovic04}
Krajnoviϙʤ, D., Cappellari, M., Emsellem, E., McDermid, R., de Zeeuw, P. T., 2004, \mnras, 357, 1113.

\bibitem[Kronawitter et al.(2000)]{kronawitter00}
Kronawitter, A., Saglia, R. P., Gerhard, O., Bender, R., 2000, $A\&AS$, 144,53.

\bibitem[Lucy(1974)]{lucy}
Lucy, L. B., 1974, \aj, 79, 745. 

\bibitem[Magorrian(1999)]{magorrian}
Magorrian, J., 1999, \mnras, 302, 530.
 
\bibitem[Magorrian et al.(1998)]{magog98}
Magorrian, J., Tremaine, S., Richstone, D., Bender, R., Bower, G., Dressler, A., Faber, S. M., Gebhardt, K., Green, R., Grillmair, C., 1998, \aj, 115, 2285.

\bibitem[Mengersen \& Tweedie(1996)]{mengersen}
Mengersen, K.L.., Tweedie, R., L., 1996, {\em {Ann. Statist.}}, 24, 101-121.

\bibitem[Merritt, Meylan \& Mayor(1997)]{merrittmeylanmayor}
Merritt, D., Meylan, G., \& Mayor, M., 1997, \aj, 114, 1074.

\bibitem[Merritt \& Tremblay(1993)]{merritttremblay}
Merritt, D., \& Tremblay, B., 1993, \aj, 106, 2229.

\bibitem[Metropolis et. al(1953)]{metropolis}
Metropolis, N., Rosenbluth, A. W., Rosenbluth, M. N., Teller, A., $\&$ Te\
ller, H., 1953, {\em Jl. of Chemical Physics}, {21}, 1087.

\bibitem[Palmer(1994)]{palmer}
Palmer, P. L., 1994, \mnras, 266, 697.

\bibitem[Richardson(1972)]{richardson}
Richardson, W.H., 1972, $Jl.\: of\: Optical\: Society\: America$, 62, 55

\bibitem[Roberts, Gelman \& Gilks(1997)]{roberts_97a}
Roberts, G., A.~Gelman, and W.~Gilks, 1997,
{\em The Annals of Applied Probability}, 7, 110.

\bibitem[Roberts and Sahu(1997)]{roberts_97b}
Roberts, G. and S.~Sahu, 1997,
{\em Journal of the Royal Statistical Society. Series B}, 59, 291.

\bibitem[Roberts \& Rosenthal(2001)]{roberts_01}
Roberts, G.~O. and J.~S. Rosenthal, 2001,
{\em Statistical Science}, 16 (4), 351--367.

\bibitem[Romanowsky \& Kochanek(1997)]{romkoch}
Romanowsky, A. J., \& Kochanek, C. S., 1997, \mnras, 287, 35, RK.

\bibitem[Rybicki(1987)]{rybicki87}
Rybicki, G. B., 1987, $IAUS$, 127, 397. 

\bibitem[Sereno et al.(2006)]{sereno_06} 
Sereno, M., De Filippis, E., Longo, G., \& Bautz, M.~W.\ 2006, \apj, 645, 170 

\bibitem[Simonneau, Varela $\&$ Munoz-Tunon(1998)]{simonneau98}
Simonneau, E., Varela, A. M., Munoz-Tunon, C., 1998, Il Nuovo
Cimento, Vol 113 B, 927.

\bibitem[Strom et al.(1981)]{strom81}
Strom, S. E., Strom, K. M., Wells, D. C., Forte, J. C., Smith, M. G., Harris, W. E., 1981, \apj, 245, 416.

\bibitem[Sha $\&$ Saul(2005)]{sha_05}
Sha, F., Saul, L., K., 2005, {\emph{Proceedings of the Twenty Second Internati\
onal Conference on Machine Learning (ICML-05), Bonn, Germany}}, 785.

\bibitem[Sun et. al(2006)]{sun_06}
Sun, J., Boyd, S., Xiao, L., Diaconis, P., 2006, {\emph{SIAM Review}}, 48, 681.

\bibitem[Tanner(1996)]{tanner} Tanner, M. A., 1996, 
{\emph{Tools for statistical inference}}, Springer-Verlag, New York.

\bibitem[Tierney(1994)]{tierney} Tierney, L., 1994, {\emph{The Annals
      of Statistics}}, 22, 1701-1762.

\bibitem[Titterington(1988)]{titterington} Titterington D.M. 1988,
\emph{IMS Lecture Notes--Monograph Series, ed. Possolo, A., Volume 20}
(Hayward, CA: Institute of Mathematical Statistics, 1991), 462.

\bibitem[Thomson et al.(1991)]{thomson_91}
Thompson, A.M., Brown, J.C., Kay J.W., Titterington D.M., 1991, \emph{IEEE Transactions to Patten Analysis $\&$ Machine Intelligence}, 13, 326. 

\bibitem[Weinberger $\&$ Saul(2006)]{weinberger_06} Weinberger, K.,
  Q., Saul, L., K., 2006, \emph{Proceedings of the Twenty First
    National Conference on Artificial Intelligence (AAAI-06), Boston,
    USA}.

\bibitem[Wang(2006)]{wang_06} Wang, A., Cherry, C., Lizotte, D.,
  Schuurmans, D., 2006, \emph{Proceedings of the 10th Conference on Computational Natural Language Learning (CONLL), NY, USA}

\bibitem[Yong et al.(2006)]{yong_06} 
Yong, K., Sahoo, Y., Roy Choudhury, K., Swihart, M. T., Minter, J. R. Prasad, P. N., 2006, {\em Nano Letters}, 6, 709.



\end{thebibliography}
\end{document}